\documentclass[twocolumn]{aastex63}
\usepackage{amsmath}

\def \kms {km\,s$^{-1}$\,}
\def \mas {mas\,yr$^{-1}$\,} 
\graphicspath{{./}{figures/}}
\usepackage{longtable}

\begin{document}

\title{HST Proper Motion Measurements of Supernova Remnant N132D: Center of Expansion and Age}

\correspondingauthor{John Banovetz}
\email{jbanovet@purdue.edu}

\author[0000-0003-0776-8859]{John Banovetz}
\affil{Department of Physics and Astronomy, Purdue University, 525 Northwestern Avenue, West Lafayette, IN 47907, USA}
\affil{Brookhaven National Laboratory, Upton, New York, United States}

\author[0000-0002-0763-3885]{Dan Milisavljevic}
\affil{Department of Physics and Astronomy, Purdue University, 525 Northwestern Avenue, West Lafayette, IN 47907, USA}
\affiliation{Integrative Data Science Initiative, Purdue University, West Lafayette, IN 47907, USA}

\author{Niharika Sravan}
\affil{California Institute of Technology, 1200 E.\ California Blvd, Pasadena, CA 91125, USA}

\author[00000-0002-4471-9960]{Kathryn E.\ Weil}
\affil{Department of Physics and Astronomy, Purdue University, 525 Northwestern Avenue, West Lafayette, IN 47907, USA}

\author[0000-0001-8073-8731]{Bhagya Subrayan}
\affil{Department of Physics and Astronomy, Purdue University, 525 Northwestern Avenue, West Lafayette, IN 47907, USA}

\author[0000-0003-3829-2056]{Robert A.\ Fesen}
\affil{Department of Physics and Astronomy, 6127 Wilder Laboratory, Dartmouth College, Hanover, NH 03755, USA}

\author[0000-0002-7507-8115]{Daniel J.\ Patnaude}
\affil{Center for Astrophysics \textbar\  Harvard \& Smithsonian, 60 Garden Street, Cambridge, MA 02138, USA}

\author[0000-0002-7507-8115]{Paul P.\ Plucinsky}
\affil{Center for Astrophysics \textbar\ Harvard \& Smithsonian, 60 Garden Street, Cambridge, MA 02138, USA}

\author[0000-0003-1413-1776]{Charles J.\ Law}
\affil{Center for Astrophysics \textbar\ Harvard \& Smithsonian, 60 Garden Street, Cambridge, MA 02138, USA}

\author[0000-0003-2379-6518]{William P.\ Blair}
\affil{The William H. Miller III Department of Physics and Astronomy, 
Johns Hopkins University, 3400 N. Charles Street, Baltimore, MD, 21218}

\author{Jon A.\ Morse}
\affil{6 BoldlyGo Institute, 31 W 34 St, Floor 7 Suite 7159, New York, NY 10001, USA}
\affil{Visiting Associate in Astronomy, Division of Physics, Mathematics and Astronomy, California Institute of Technology, Pasadena, CA 91125}

\begin{abstract}
We present proper motion measurements of oxygen-rich ejecta of the LMC supernova remnant N132D using two epochs of Hubble Space Telescope Advanced Camera for Surveys data spanning 16 years. The proper motions of 120 individual knots of oxygen-rich gas were measured and used to calculate a center of expansion (CoE) of $\alpha$=$5^{h}25^{m}01.71^{s}$
and $\delta$=$-69^{\circ}38^{\prime}41\farcs64$ (J2000) with a 1-$\sigma$ uncertainty of $2\farcs90$. This new CoE measurement is $9\farcs2$ and $10\farcs8$ from two previous CoE estimates based on the geometry of the optically emitting ejecta. We also derive an explosion age of 2770 $\pm$ 500 yr, which is consistent with recent age estimates of $\approx 2500$ yr made from 3D ejecta reconstructions.  We verify our estimates of the CoE and age using a new automated procedure that detected and tracked the proper motions of 137 knots, with 73 knots that overlap with the visually identified knots. We find the proper motions of ejecta are still ballistic, despite the remnant's age, and are consistent with the notion that the ejecta are expanding into an ISM cavity. Evidence for explosion asymmetry from the parent supernova is also observed. Using the visually measured proper motion measurements and corresponding center of expansion and age, we compare N132D to other supernova remnants with proper motion ejecta studies.

\end{abstract}


\section{Introduction} \label{sec:intro}

\begin{figure*}[!htp]
\centering
\includegraphics[width=0.95\textwidth, angle=0]{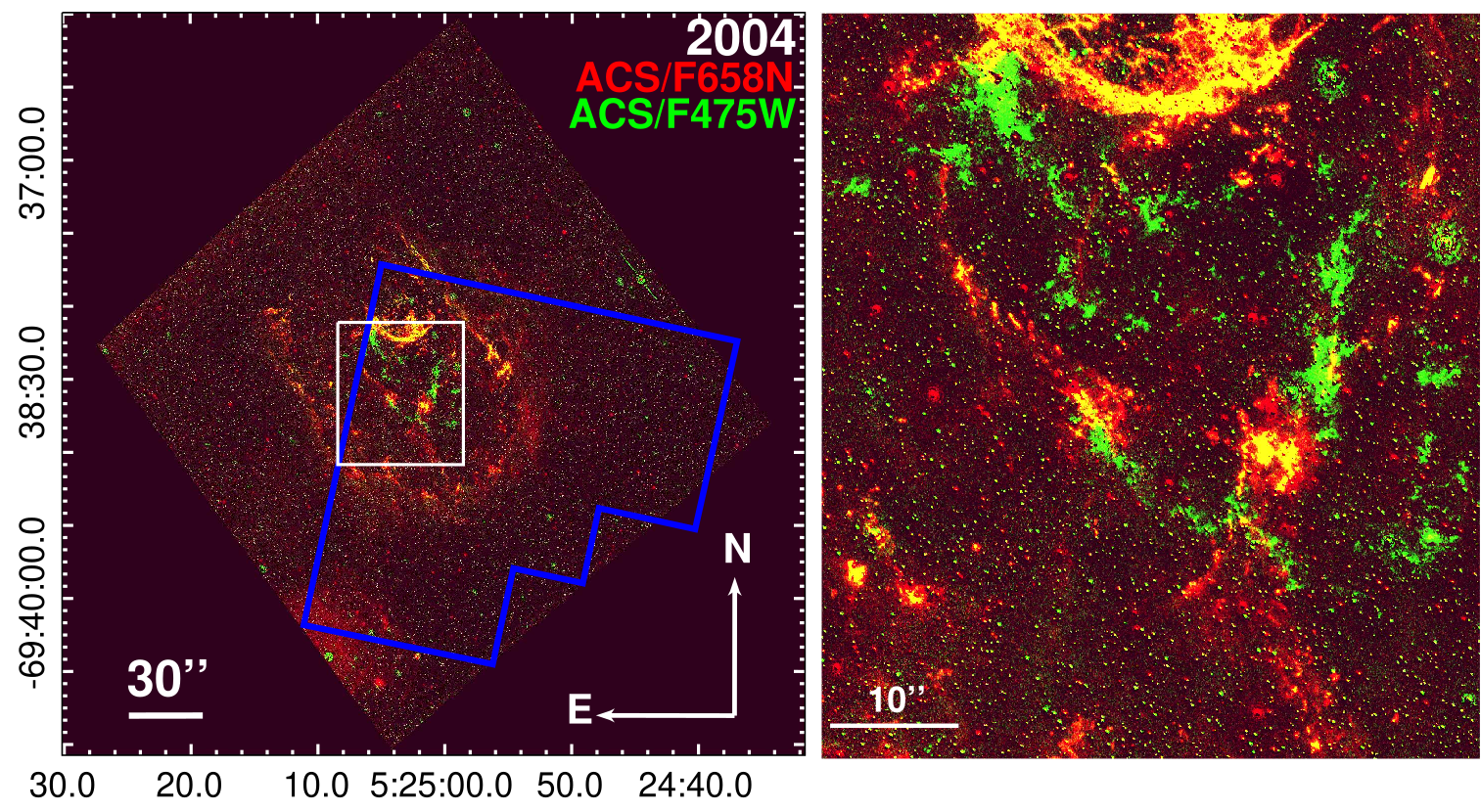}
\caption{Left: Continuum subtracted image of N132D using images taken with HST/ACS and the F475W (green) and F658N (red) filters (additional image information can be found in Table \ref{tab:images}). 
The blue polygon indicates the field of view of the 1994 WFPC2/F502N image. The white square is the cropped image of N132D used for proper motion measurements.
Right: Enlarged view of the white square, highlighting in green the O-rich ejecta used for the proper motion measurements.
}
\label{fig:geomapref}
\end{figure*}

Supernova remnants (SNRs) provide valuable insights into the explosion processes of supernovae that are otherwise too distant to resolve \citep[see][for a review]{Milisavljevic2017}. They offer unique opportunities to probe the elemental distribution of metal-rich ejecta and investigate the progenitor star's mass loss history at fine scales \citep[see][for a review]{Lopez2018}. Young, nearby oxygen-rich (O-rich) SNRs, created from the collapse of massive stars \citep[ZAMS mass $> 8M_{\odot}$;][]{Smartt2009}, 
are especially informative to study core-collapse dynamics because they are often associated with progenitor stars that were largely stripped of their hydrogen envelopes \citep[e.g.,][]{Blair2000,Chevalier05,Temim2022}.  The kinematic and chemical properties of their metal-rich ejecta retain information about the parent supernova explosion that would otherwise be lost in an H-rich explosion  \citep{Milisavljevic2010}.  

Tracking metal-rich ejecta over many years and measuring their proper motion enables estimates of the center of expansion (CoE) and explosion age, as well as information about the progenitor system's circumstellar material (CSM) environment via ejecta interaction. The CoE and explosion age are important values for determining the kick velocity of compact objects \citep{Vogt2018,Banovetz2021,Long2022}, searching for surviving companions \citep{Kerzendorf2019,Li2021}, and measuring differences between optical and X-ray centers \citep{Katsuda2018}. These values can also serve as important tests for increasingly sophisticated 2D and 3D supernova simulations \citep[e.g.,][]{Wongwathanarat2015,Janka2016,Burrows2019,Ferrand2021,Orlando2021,Orlando2022}.

Only a handful of known O-rich SNRs are sufficiently resolved to measure proper motion of high velocity ejecta from multi-epoch observations. This small list includes Cassiopeia A \citep[Cas A;][]{Kamper1976,Thorstensen2001,Fesen2006,Hammell2008}, G292.0+1.8 \citep[G292;][]{Murdin1979,Winkler2009}, and 1E 0102.2-7219 \citep[E0102;][]{Finkelstein2006,Banovetz2021}. This paper focuses on the O-rich SNR N132D, which to date has no published proper motion measurements of its optically-emitting ejecta.

N132D is located in the bar of the Large Magellanic Cloud (LMC) and was first identified as a SNR from radio emission \citep{West1966}. Later, it was found to contain high velocity O-rich ejecta through optical spectra, classifying it as an O-rich SNR \citep{Danziger1976,Lasker1980}. The parent supernova may have been a Type Ib with a 10-35 $M_{\odot}$ ZAMS progenitor \citep{Blair2000,Sharda2020}. Presently, the supernova continues to expand into a cavity created by the pre-supernova mass loss of the progenitor star \citep{Hughes1987,Sutherland1995,Blair2000,Chen2003,Sharda2020}.

N132D is the brightest X-ray and gamma-ray SNR in the LMC \citep{Clark1982,Favata1997,Borkowski2007,HESS2015,Ackermann2016}. X-ray images show a horseshoe shaped forward shock \citep[e.g.,][]{Borkowski2007,Bamba2018}, the southern portion of which is associated with natal molecular clouds \citep{Banas1997,Dopita2018,Sano2020}.
X-ray and radio observations indicate that N132D is transitioning from a young to middle-aged SNR and is about to enter the Sedov phase \citep{Dickel1995,Favata1997,Bamba2018}.

\begin{deluxetable*}{lccccccc}[!thb]
\label{tab:images}
\tablecaption{HST Observations of N132D}
\tablehead{
\colhead{PI} & \colhead{Date} & \colhead{Exp.\ Time} & \colhead{Instrument} & \colhead{Filter} & \colhead{$\lambda_{\rm center}$} & \colhead{Bandwidth} & \colhead{Pixel Scale}\\
 & & (s) & & & (\r{A}) & (\r{A}) & ($^{\prime\prime}$ pixel$^{-1}$)\
}
\startdata
Blair & 1994/08/09 & 3600 & WFPC2/PC & F502N & 5012 & 27 & 0.0455\\
Green & 2004/01/22 & 1440 & ACS/WFC & F658N & 6584 & 75 & 0.049\\
Green & 2004/01/22 & 1800 & ACS/WFC & F550M & 5580 & 389 & 0.049\\
Green & 2004/01/21 & 1440 & ACS/WFC & F775W & 7702 & 1300 & 0.049\\
Green* & 2004/01/22 & 1520 & ACS/WFC & F475W & 4760 & 1458 & 0.049\\
Milisavljevic* & 2020/01/05 & 2320 & ACS/WFC & F475W & 4760 & 1458 & 0.049\\
Milisavljevic & 2020/01/05 & 2480 & WFC3/UVIS & F502N & 5013 & 48 & 0.040\\
\enddata
\tablecomments{* denotes images used in proper motion analysis}
\end{deluxetable*}

Previous estimates of N132D's explosion age have been made by dividing the radius of the SNR by the maximum radial velocity of the ejecta, 
yielding age estimates ranging from 1300-3440 yr \citep{Danziger1976,Lasker1980,Morse1995,Sutherland1995}. \cite{Morse1995} gave two estimates for the CoE of N132D. The first estimate was made by fitting an ellipse to the diffuse outer rim, and the second by finding the geometric center of O-rich ejecta. Recent 3D reconstructions of N132D use this geometrically-derived center as the CoE and find that N132D's optically emitting oxygen-rich material is arranged in a torus distribution, inclined at an angle of $\approx25-28^{\circ}$ in the plane of the sky \citep{Vogt2011,Law2020}. They also provide the most recent age estimates of $\approx2500$ yr.  

This paper uses high resolution images obtained with the Hubble Space Telescope (HST) to measure proper motion of N132D's oxygen-rich ejecta to estimate a CoE and explosion age. Section \ref{sec:obs} discusses observations of N132D and the images used. Section \ref{sec:Manual} describes our proper motion measurements and analysis techniques. Section \ref{sec:Automated} introduces an automated procedure to measure proper motions using computer vision. Section \ref{sec:discussion} discusses the implications of the proper motion measurements, CoE, and explosion age as it pertains to previous estimates and other SNRs. We summarize and conclude in Section \ref{sec:conclusion}.

\begin{figure}[tp]
\centering
\includegraphics[width=0.95\linewidth, angle=0]{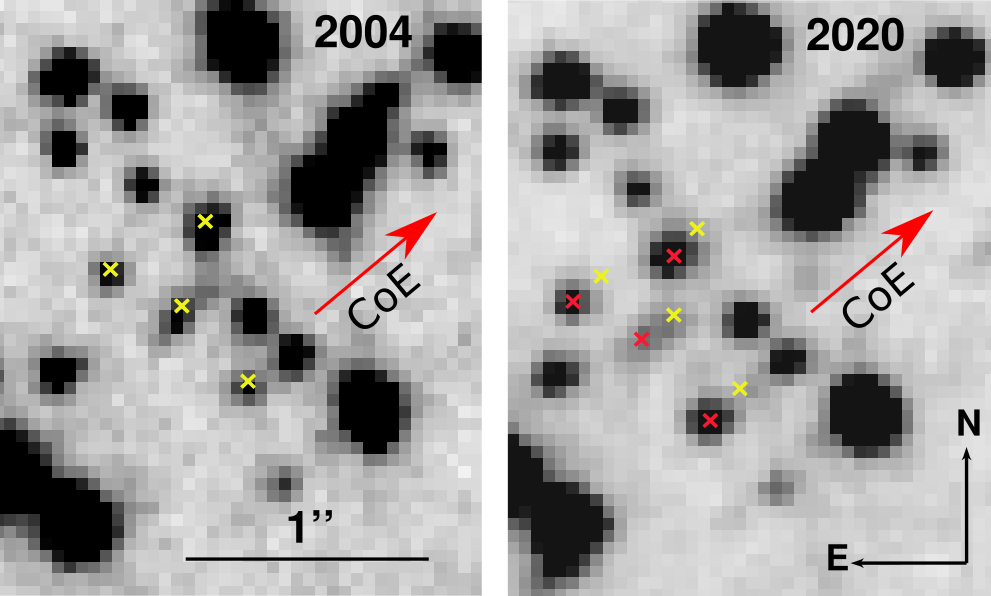}

\caption{ACS/F475W images showing examples of the expanding ejecta knots in 2004 (left) and 2020 (right) in the area of the runaway knot (RK, see Section \ref{sec:PM_Discussion} for more details). The 2004 knot centroids are shown as yellow crosses while the 2020 centroids are shown as red crosses. The red arrow points in the direction of the CoE.}
\label{fig:Tracking}
\end{figure}

\section{Observations} \label{sec:obs}

Using the Mikulski Archive for Space Telescopes (MAST) at the Space Telescope Science Institute, we examined three epochs of HST images that are sensitive to [O III] $\lambda\lambda$4959, 5007 emission tracing oxygen-rich ejecta of N132D\footnote{The specific observations analyzed can be accessed via \dataset[10.17909/4ppy-4e90]{https://doi.org/10.17909/4ppy-4e90}}. These consist of  an image taken in 1994 using the Wide Field Planetary Camera 2 (WFPC2) with the F502N filter (PI: Blair GO-5365), a 2004 image using the the Advanced Camera for Surveys (ACS) and the F475W filter (PI: Green GO-12001), and a 2020 image using the ACS/F475W setup (PI: Milisavljevic GO-15818).  The 1994 and 2020 F502N images use different instrument and filter configurations, whereas the 2004 and 2020 F475W images were both obtained with ACS. Utilizing the same camera/filter setup greatly improves the tracking confidence of the gas, as using different camera/filters setups can cause ambiguity in precise tracking due to brightening effects \citep[see][]{Banovetz2021}. Thus, only the ACS images were used for proper motion tracking. The 2020 F502N image was used to confirm O-rich ejecta emission from possible continuum emission. All images were processed using \textit{Astrodrizzle} \citep{Gonzaga2012} and had a final image scale of approximately $0\farcs05$ pixel$^{-1}$. Table \ref{tab:images} contains more information about the images used for analysis.

To align the images, we use the \texttt{geomap} task in  PYRAF\footnote{PYRAF is distributed by the National Optical Astronomy Observatory, which is operated by the AURA, Inc., under cooperative agreement with the National Science Foundation. The Space Telescope Science Data Analysis System (STSDAS) is distributed by STScI.} 
to create a transformation database using 30 anchor stars between the two images (see Table \ref{tab:stars} in Appendix). These anchors were chosen for their low proper motions and small transformation residuals. The transformation had resulting residuals of $\approx0.3$ pixels ($\approx0\farcs015$). 
We then used the PYRAF task \texttt{geotran} to apply this transformation, aligning the images. Once the images were aligned, they were cropped to a $51\farcs5\times 58\farcs3$ field of view that contains only the O-rich portion of the remnant (see Figure \ref{fig:geomapref}). The cropping extent was determined by visual examination of oxygen emission, and we ensured that all high proper motion ejecta knots were contained within the selected field of view.  The World Coordinate System (WCS) was calculated using a locally compiled version of the  \texttt{Astrometry.net\footnote{Astrometry is distributed as open source under the GNU General Public License and was developed on Linux.}} \citep{Lang2010}. This WCS solution is accurate to $\approx0\farcs17$ and was taken into account for the final CoE error.

\section{Proper Motion Measurements: Manual Estimation}
\label{sec:Manual}

\begin{figure}[!htb]
\centering
\includegraphics[width=1\linewidth]{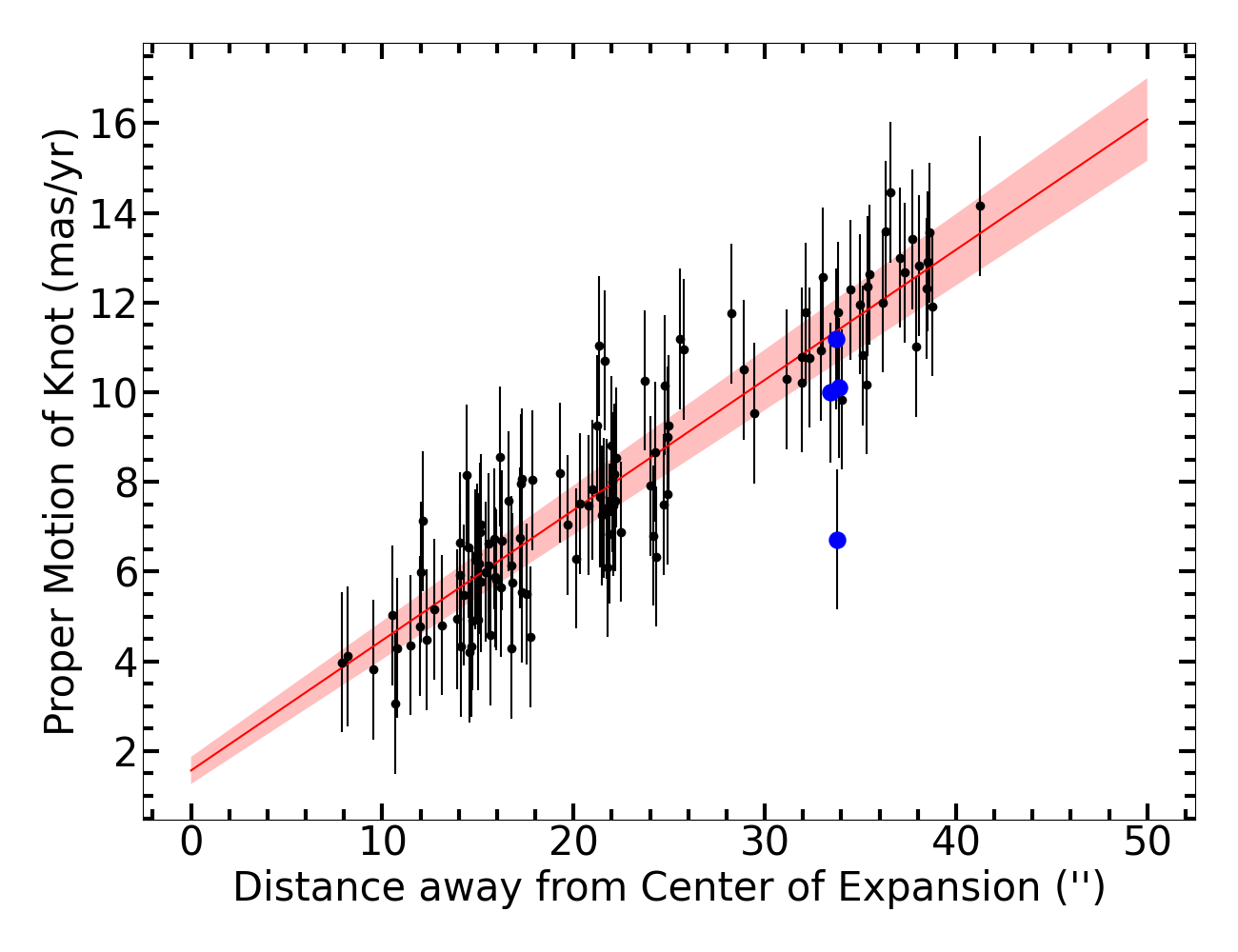} 

\caption{The absolute proper motion vs radial distance of the knots. The proper motions of the knots of ejecta are shown as black points with their corresponding 1-$\sigma$ error. The red line indicates a linear fit to the data with the shaded region indicating the 1-$\sigma$ error. The location of the four knots in the RK region are highlighted as blue points.}

\label{fig:V_Radial}
\end{figure}

\begin{figure*}[tp]
\centering
\includegraphics[width=0.7\linewidth, angle=0]{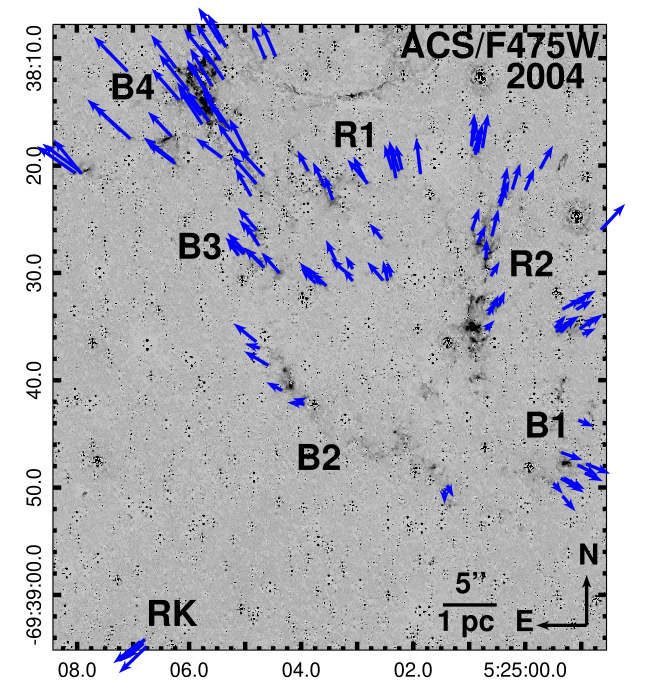}
\caption{2004 ACS/F475W continuum- and hydrogen-subtracted image with vectors representing the measured shifts (multiplied by a factor of 20 for visual clarity) shown in blue. The regions identified in \citet{Morse1995} are also labeled.
} 

\label{fig:vector}
\end{figure*}

Using the aligned images, we identified knots with high proper motions. Knots were chosen by how well they could be tracked visually, with most knots being optically bright and circular to have confidence in the measurements. The shifts of the knots were calculated by blinking between the two images in SAOImage DS9 and visually locating the centers of the knots or other conspicuous features (see Figure \ref{fig:Tracking}). The centers were measured multiple times to estimate positional errors of each knot. During the 16 yrs, knots can possibly brighten/dim or change morphology as they interact with the surrounding medium \citep{Fesen2011,Banovetz2021}. This interaction can skew results if using the astrometric approach of fitting a Gaussian for the knots. We did apply a Gaussian based centroid fitting procedure (see Appendix and Section \ref{sec:AutovVisual}) and found more accurate results through manual inspection.

We applied our methodology to 120 knots (see Table \ref{tab:knots} in Appendix\footnote{Also available in a machine readable format}), which resulted in proper motions ranging from 3--14.5 miliarcseconds (mas) per year, with a median proper motion of 7.54 \mas and average relative error of $13\%$ (Figure \ref{fig:V_Radial}). 
This translates to a median velocity of 1784 \kms assuming a distance to the LMC of 50 kpc \citep{Panagia1991}. Our median velocity is consistent within uncertainties to the average expansion velocity of 1745 \kms calculated using a fitted projected radius of N132D \citep{Law2020}. The linear fit also gives a higher scaling factor $S$ of $0\farcs014$ per \kms compared to $0\farcs010$ per \kms of \citet{Law2020}. Figure \ref{fig:vector} shows the locations and proper motions of the 120 knots, as well as the O-rich regions discussed in \citet{Morse1995}.

\subsection{Center of Expansion} \label{sec:CoE}

Our approach to determine the CoE of N132D uses the trajectories of the ejecta augmented with a likelihood function. This method is similar to that used by \cite{Banovetz2021} and \cite{Thorstensen2001} for the calculation of E0102's and Cas A's CoE, respectively. We favor this method because it only depends on the direction of the knots, and is not sensitive to deceleration over time. 

We assume that the likelihood of the CoE in the plane of sky coordinates (X,Y) is given by:
\begin{equation}
    \mathcal{L}(X,Y)=\Pi_{i}\frac{w_{i}}{2\sigma_{i}}exp(-d^{2}_{i\perp}/(2\sigma_{i}^{2}))
\label{eq:CoE}
\end{equation}
\begin{equation}
    w_{i}=\frac{1}{P_{y_{i}}P_{x_{i}}},
\label{eq:Weight}
\end{equation}

\noindent where $d_{i \perp}$ is the perpendicular distance between (X,Y) and the knot's line of position, and $\sigma_{i}$ is the uncertainty associated with the point common to the knot's extended line of position and $d_{i \perp}$ \citep{Banovetz2021}. We also define $w$, the probability of finding an individual knot in a given X and Y position, denoted by $P_{x_{i}}$ and $P_{y_{i}}$, respectively, which was calculated using a kernel density estimate (KDE) to fit knots in the (X,Y) plane. 
The (X,Y) combination that maximizes this function gives the CoE. 
The uncertainty of the CoE is derived from 100,000 artificial data sets generated from position and direction distributions of individual knots. 

A notable difference from \citet{Banovetz2021} is the addition of a weight, $w$. We used this weight to minimize the effects of selection bias in our sample. As seen in Figure \ref{fig:vector}, N132D is unique compared to other O-rich SNRs in that the knot distribution is skewed, with a larger number of knots displaying proper motions in the northern region of the remnant as compared to the southern region. Without the weight, the CoE will skew in the direction of the more populated region. This added weight term compensates for sparse regions by giving proportionally more weight to knots in these regions.

\begin{figure*}[tp]
\centering
\includegraphics[width=0.45\textwidth, angle=0]{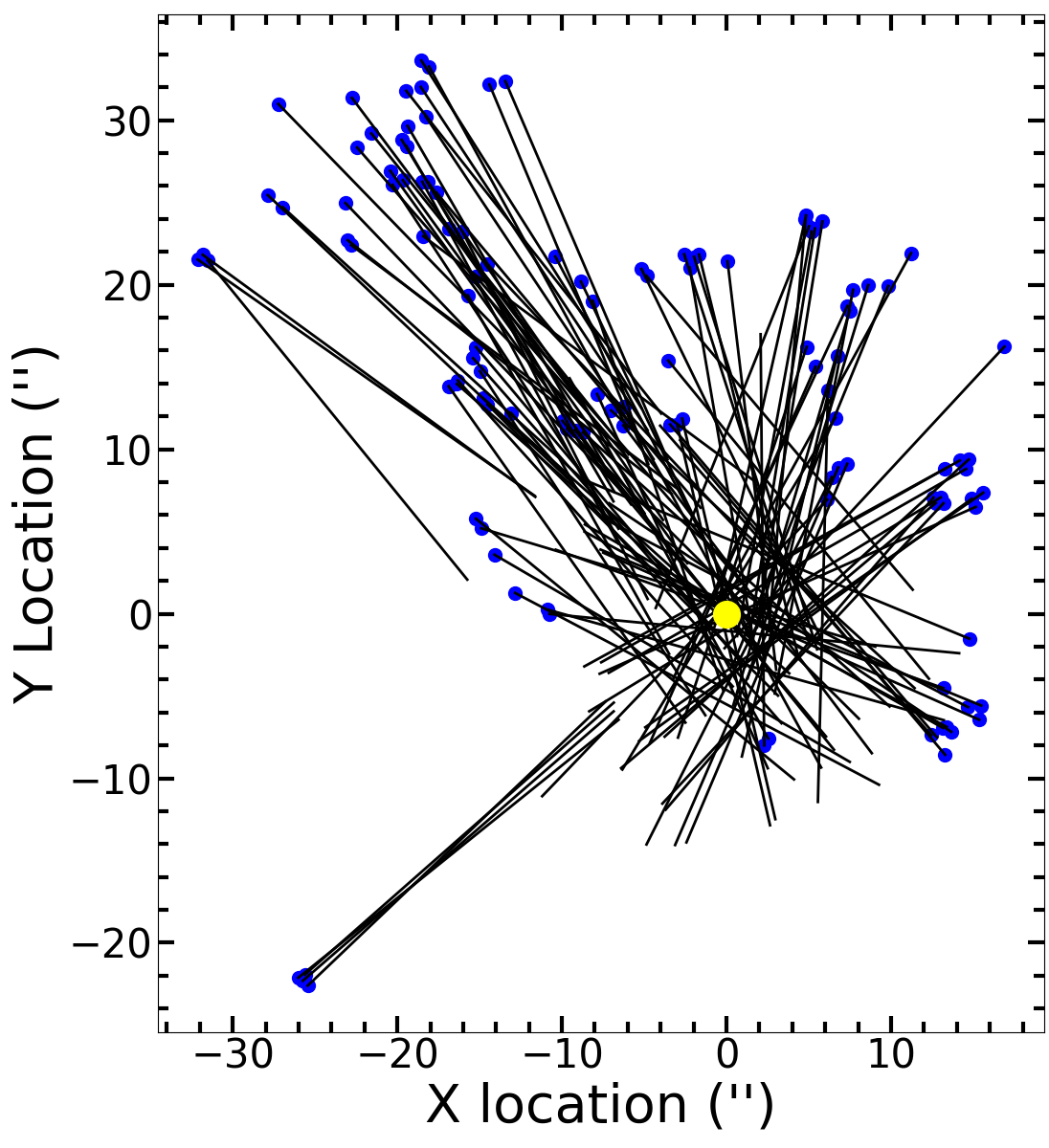}
\includegraphics[width=0.45\textwidth, angle=0]{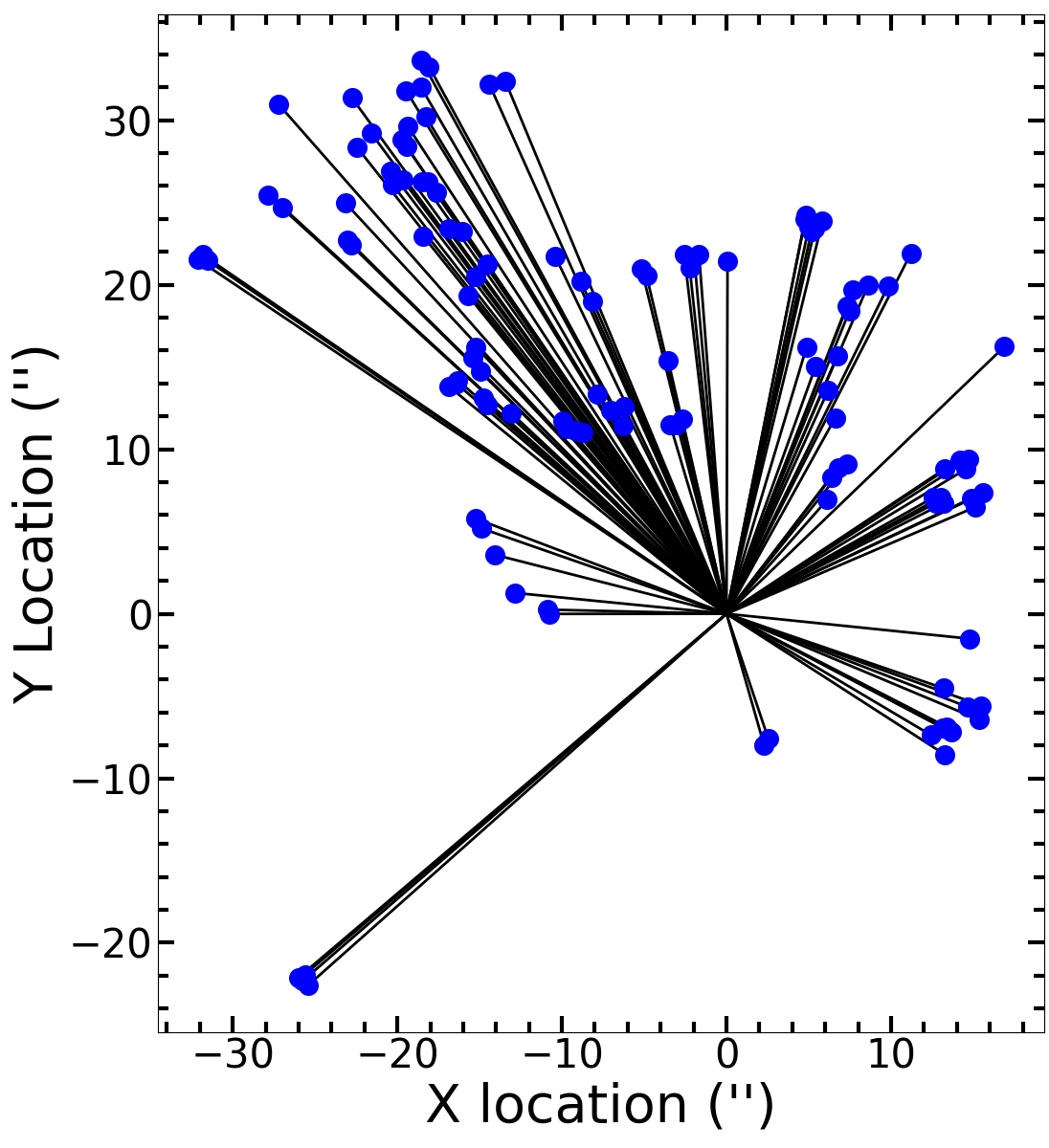}
\caption{Left: The visually measured proper motions of the 120 knots traced back 25$^{\prime\prime}$ ($\approx$3000 yr assuming an average proper motion of 8.1 \mas). The CoE is shown in yellow. Right: The trajectories of the visually measured knots if forced to originate from our calculated CoE. Strong spatial asymmetry in the knot distribution is observed with respect to the CoE.} 
     
\label{fig:Fesen}
\end{figure*}

Applying this procedure to our proper motion measurements
yields a CoE of
$\alpha$=5$^{h}$25$^{m}$01.71$^{s}$
and $\delta$=-69$^{\circ}$38$^{\prime}$$41\farcs64$ (J2000) with a 1-$\sigma$ uncertainty of $2\farcs90$. 
Figure \ref{fig:Fesen} shows the trajectories of the knots as compared to the derived CoE. 

\subsection{Explosion Age} \label{sec:ExpAge}

\begin{figure}[!htb]
\centering
\includegraphics[width=0.47\textwidth, angle=0]{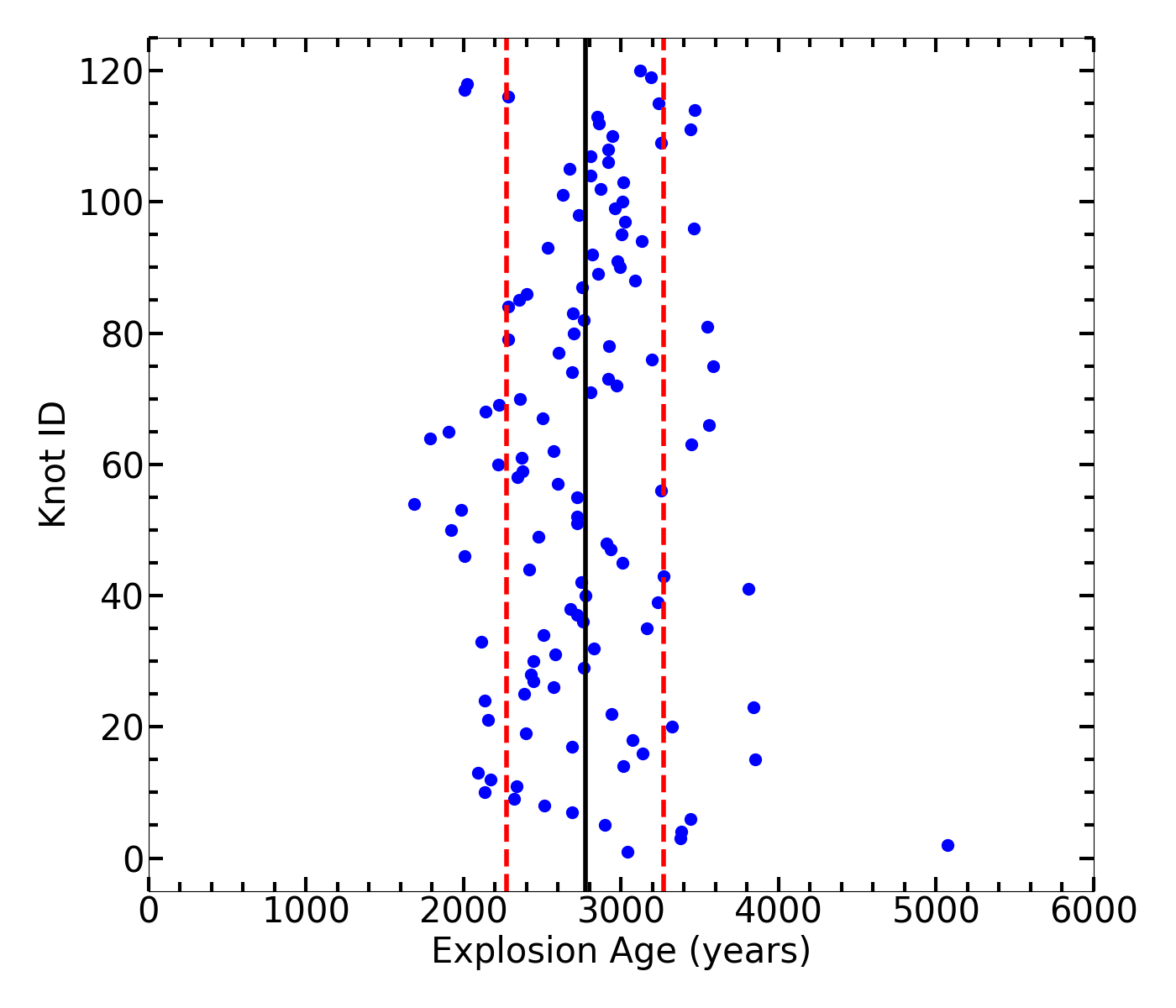}
\caption{A comparison of the explosions age measurements, assuming our CoE. The black line represents the average age of the data set, while the red dashed line is the 1-$\sigma$ uncertainty. This results in an explosion age of 2770 $\pm$ 500 yr. 
}
\label{fig:Fesenyears}
\end{figure}

Using the manually tracked knots and the associated center of expansion estimate, we calculated the explosion age of N132D by dividing the knot's distance from the CoE by their proper motion measurements. Figure \ref{fig:Fesenyears} shows the calculated explosion age of all 120 knots. Combining these ages resulted in an age of 2770 $\pm$ 500  yr.

We also calculated the explosion age using only the knots with the fastest proper motions. A similar approach was used by \citet{Fesen2006} for the explosion age of Cassiopeia~A and \citet{Banovetz2021} for E0102. This method assumes that knots with the fastest proper motions are least decelerated, 
resulting in a more accurate explosion age. Forty-nine of the 120 knots with proper motions greater than the average (8.1 \mas) were selected. 
Almost all these knots correspond to the region B4 from \citet{Morse1995}. Using these knots resulted in an explosion age of 2745 $\pm$ 404 yr, consistent with the age using all of the knots. For further discussion, we adopt the age of 2770 $\pm$ 500  yr, as this age is representative of all the knots.

\section{Proper Motion Measurements: Automated via Computer Vision}
\label{sec:AutovVisual}

We also implemented a novel computer vision based approach to measure the proper motions of the ejecta. This approach utilizes hydrogen and continuum subtracted images between the epochs to insure that only the O-rich material is being tracked. Then, regions of high emission and/or high ejecta proper motions are specified. These regions, or stamps, are passed through an automated detection procedure to identify knots using image segmentation and deblending. To track the knots, a kernel density estimate (KDE) estimates the peaks within these segments, and we use these peaks to measure the proper motion between the epochs. A more detailed explanation of this procedure can be found in the Appendix.

The automated procedure identified and measured the proper motions of 137 knots of ejecta\footnote{Proper motion measurements and locations can be found in a machine readable format, a subset of which can be found in Table \ref{tab:knots}}. Seventy-three of these 137 knots matched visually identified knots used in the manual procedure. The average difference in inferred values between the manual and automated procedures for the shift between epochs is $0\farcs03$ (or $\approx1.8$ \mas ) and the vector angle is $\approx12^{\circ}$. The error of the proper motions was set to 0.4 pixels ($\approx0\farcs2$), from the sub-pixel ratio of the KDE. 
While the automated proper-motion measurements generally followed the same ballistic $v \propto r$ relationship found with the visually tracked knots, the measurements exhibited more scatter and higher uncertainty. This discrepancy most likely arises from tracking fainter, less dense knots that are more susceptible to deceleration compared to the bright, denser knots that were found visually.

While our automated procedure generally produces similar results to the visually measured proper motions, the sample is contaminated by the less dense knots, possibly skewing the results. Hence, we adopt the visual measurement results for this paper. We note that, the results above used conservative metrics in the measurement of the proper motions in order to not be biased too heavily by any one parameter. Attempts to improve the results by fine-tuning these parameters can be found in the Appendix.

\begin{figure*}[tp]
\centering
\includegraphics[width=0.98\linewidth, angle=0]{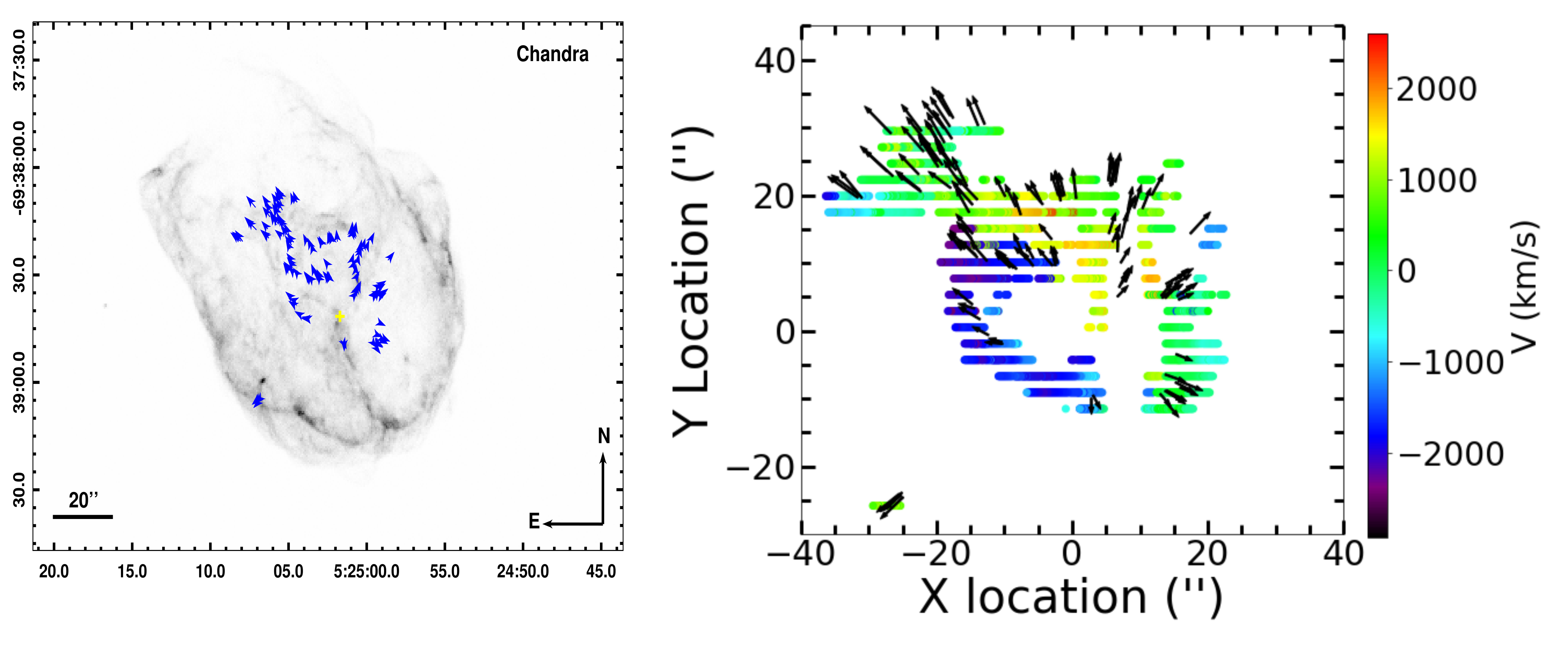}
\caption{Left: Chandra image of N132D (PI: Borkowski). The proper motions from the visual measurements are in blue and our calculated CoE in yellow (see Section \ref{sec:CoE}). Right: Vectors of proper motions (black) with Doppler velocities \citep{Law2020}, centered on our CoE (see Section \ref{sec:CoE}).}
\label{fig:Chandra}
\end{figure*}

\begin{figure*}[tp]
\centering
\includegraphics[width=0.7\linewidth, angle=0]{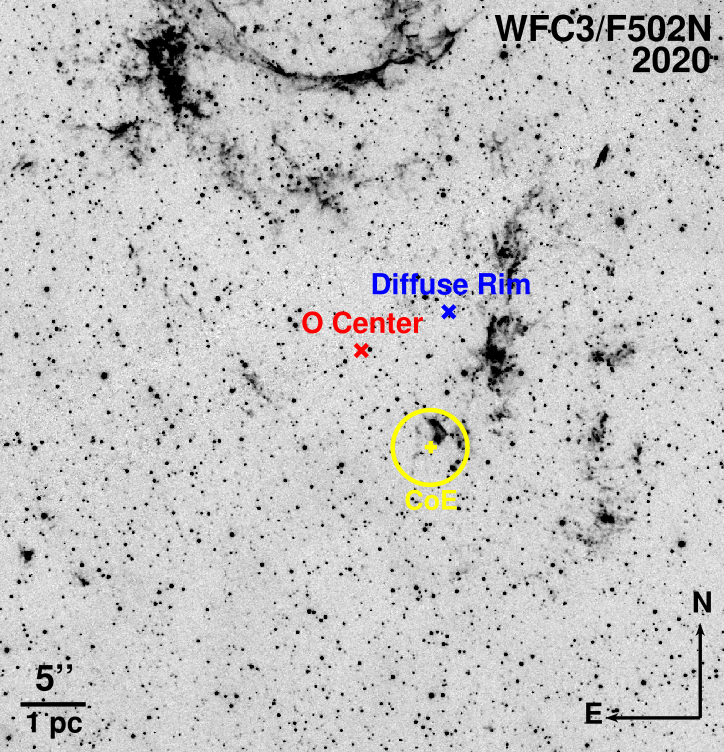}
\caption{Center of expansion estimates for N132D. This paper's CoE calculations and associated 1-$\sigma$ uncertainty is shown in yellow. This CoE is centered at $\alpha$=$5^{h}25^{m}01.71^{s}$
and $\delta$=$-69^{\circ}38^{\prime}41\farcs64$ (J2000) with a 1-$\sigma$ uncertainty of $2\farcs90$ and is $9\farcs2$ and $10\farcs8$ away from the estimates of \citet{Morse1995} found by fitting an ellipse to the diffuse outer rim (blue) and the O-rich geometric center (red).}
\label{fig:Center_of_Expansion}
\end{figure*}

\section{Discussion} 
\label{sec:discussion}

\subsection{Proper Motion Measurements}
\label{sec:PM_Discussion}

Our work presents the first proper motion measurements of the O-rich ejecta of N132D. We visually identified and tracked 120 knots of ejecta across 16 years. While the baseline is large and could be on the order of shock cooling times, we are confident in our tracking ability. This is because the emission mechanism is most likely a combination of shock excitation and photoionization producing a high amount of [O III] emission \citep{Sutherland1995}, the low densities of the shock will increase the cooling times \citep{Blair2000}, and we find similar morphology in the knots between epochs. With these 120 knots, we find that the ejecta follow homologous expansion with an average proper motion of 8.1 \mas (median proper motion of 7.54 \mas) despite N132D's advanced age approaching the Sedov phase. 

Figure \ref{fig:V_Radial} shows the proper motions of the knots versus their distance away from our calculated CoE (see Section \ref{sec:CoE}). Comparing this with spectroscopic measurements, the highest proper motion measurements are seen in the B4 region, as first reported in \citet{Morse1995}, and shown in the left panel of Figure \ref{fig:vector}. This is to be expected, as B4 corresponds to a region of small Doppler velocities in the O-rich ejecta \citep{Morse1995,Vogt2011,Law2020}, as seen in the upper left of the plot in the right panel of Figure \ref{fig:Chandra}. We find this inverse relationship between the proper motion measurements and Doppler velocities to hold true except for the region B1. B1 also corresponds to an area of small Doppler velocities, but the proper motion measurements are much smaller compared to B4. 
This difference in proper motions could be a result of B1 possibly being outside the reverse shock, as proposed by \cite{Vogt2011} (see Figure \ref{fig:Chandra} for X-ray emission tracing the shocks).
However, as this region is consistent with the ballistic trend, it is more likely associated with an explosion asymmetry (see Section \ref{sec:O_SNRs} for more discussion).

Notably, the fit shown in Figure \ref{fig:V_Radial} does not pass through the origin and has an offset of $\approx+1.6$ \mas. Forcing the line through the origin results in $S\approx 0\farcs012$ per \kms, which is closer to the value reported in \citet{Law2020}. This offset in the original fit could be indicative of deceleration experienced by the ejecta over time. As the ejecta expands in the surrounding environment, the fastest ejecta will interact and decelerate at a different rate compared to the slower ejecta. This will disrupt the $v \propto r$ relation between ejecta velocity and distance from the CoE, introducing a positive offset term to the linear fit.

\begin{deluxetable*}{lcccc}[tp]
\label{tab:SNR_Prop}
\tablecaption{Characteristics of Young O-rich SNRs}
\tablehead{
\colhead{Parameter} & \colhead{G292} & \colhead{E0102} & \colhead{Cas A} & \colhead{N132D}
}
\startdata
Proper motion derived CoE (J2000) & 11:24:34.4 [1] & 1:04:02.48 [2] & 23:23:27.77 [3] & 5:25:01.71$^{a}$ \\
& -59:15:51 & -72:01:53.92 & +58:48:49.4 & -69:38:41.64 \\
CoE 1-$\sigma$ Error ($^{\prime\prime}$) & 5 & 1.77 & 0.4 & 2.90 \\
Center of X-ray emission (J2000) & 11:24:33.1 [4] & 01:04:1.964 [5] & 23:23:27.9 [4] & 5:25:03.08$^{b}$ \\
 & -59:15:51.1 & -72:01:53.47 & 58:48:56.2 & -69:38:32.6 \\
Current Age (yr) &  $\sim$2990 [1]  & $\sim$1740 [2] & $\sim$350 [3] & $\sim$2770$^{a}$ \\
Distance to Remnant (kpc) & $6.2 \pm 0.9$ [7] & $62.1 \pm 1.9$ [8,9] & 3.4 $^{+0.3}_{-0.1}$ [10] & $50.1 \pm 3.1$ [11] \\
Size of Remnant (arcmin) & 8.4--9.6 [12] & 0.7 [5] & 5.6 [13] & 1.8 [14]\\
Ejecta transverse velocity (km/s) & $\sim$1500-3600 [1] & $\sim$1300--2700 [2] & $\sim$5500--14500 [15] & $\sim$700--3400$^{a}$ \\
Progenitor ZAMS Mass ($M_{\odot}$) & 13--30 [16] & 25--50 [17,18] & 15--20 [19] & 10--35 [17,20] 
\enddata
\tablenotetext{a}{This work.}
\tablenotetext{b}{Estimated using archival Chandra observations (Xi et al., private communication).}
\tablerefs{[1] \cite{Winkler2009},
[2] \cite{Banovetz2021},
[3] \cite{Thorstensen2001},
[4] \cite{Katsuda2018},
[5] \cite{Xi2019},
[6] \cite{Dickel1995},
[7] \cite{Gaensler2003},
[8] \cite{Graczyk2014},
[9] \cite{Scowcroft2016},
[10] \cite{Reed1995},
[11] \cite{Panagia1991},
[12] \cite{Park2007},
[13] \cite{Vink2022},
[14] \cite{Law2020},
[15] \cite{Fesen2006},
[16] \cite{Bhalerao2019}, 
[17] \cite{Blair2000}, 
[18] \cite{Finkelstein2006}, 
[19] \cite{Lee2014}, 
[20] \cite{Sharda2020}
}
\end{deluxetable*}

\begin{figure*}[!ht]
\centering
\includegraphics[width=0.8\textwidth, angle=0]{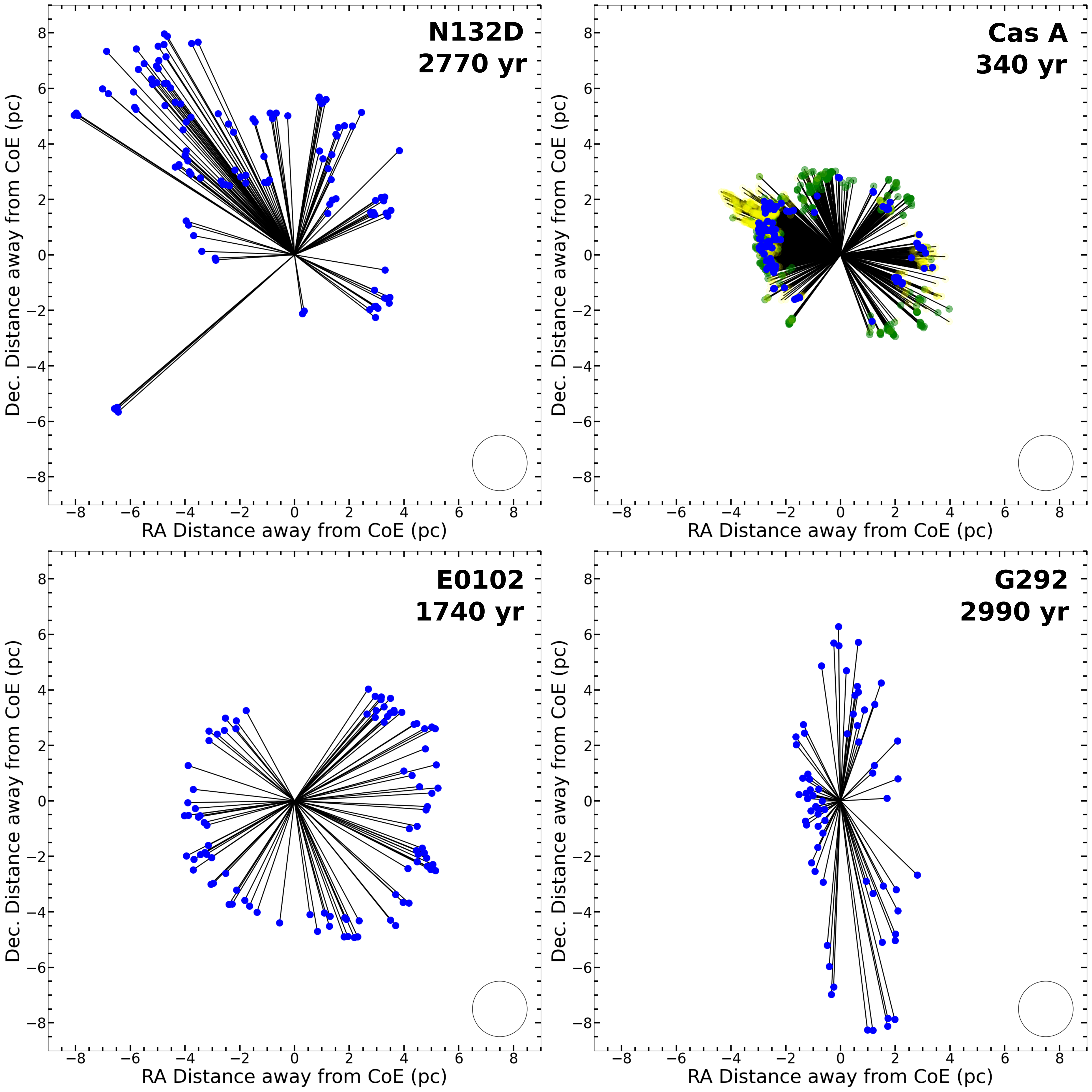}
\caption{Locations of O-rich knots (blue) from N132D (this work), E0102 \citep{Banovetz2021}, Cas A \citep{Hammell2008}, and G292 \citep{Winkler2009}, with trajectories (black) forced to originate on their proper-motion derived CoEs \citep{Thorstensen2001,Winkler2009,Banovetz2021}. Cas A also includes nitrogen rich knots (green) and fast moving knots (yellow) \citep{Hammell2008}. The circle in the bottom right is 1 pc in radius. The physical distances were calculated using  distance estimates and CoEs from Table~\ref{tab:SNR_Prop}.}

\label{fig:SNR_Comp}
\end{figure*}

A unique feature of N132D is the runaway knot (RK), which is an isolated small clump of ejecta located in the southwestern portion of the remnant and is unique in that it is enhanced in Si and S but not O \citep{Law2020}. The RK was first reported in \citet{Morse1995} and recent 3D reconstructions show that the knot is perpendicular to the main torus \citep{Vogt2011,Law2020}. Explanations for the origin of the RK include evidence of a polar jet \citep{Vogt2011} and high velocity ejecta, similar to Cas A \citep{Fesen1996,Law2020}. We were able to find and measure the proper motions of four O-rich knots in the same region as the RK (see Figure \ref{fig:Tracking} and Figure \ref{fig:Auto_procedure}). The proper motions are faster than the global average but are still consistent with proper motions of other knots ($\approx9.5$ \mas or $\approx2250$ \kms ). Combining this with a Doppler velocity of 820 \kms \citep{Law2020}, the RK has a 3D velocity of $\approx 2395$ \kms . This is much lower than the total spatial velocity of $\approx 3650$ \kms calculated by \citet{Law2020}, which is a consequence of using a scaling relation found from the innermost ejecta that was then extended to the RK.

Although we conclude that the RK does not have kinematic features that distinguish it from the bulk of N132D's ejecta, there remains a conspicuous gap in the proper motion measurements in the direction of the RK and our analysis was unable to uncover any new rapidly moving ejecta in this location. The isolated nature of the RK may imply that it passed through the reverse shock and only recently became optically bright again due to interaction with the CSM/ISM. This interpretation is supported by X-ray enhancement in close proximity to the RK \citep{Borkowski2007,Law2020}, which can be seen in the left panel of Figure \ref{fig:Chandra}, where the RK is located in the southeast, very close to the rim of X-ray emission. Further observations of this area may reveal more knots interacting with the CSM/ISM and fill the conspicuous gap.

\subsection{CoE and Age Results}
\label{sec:CoE_Age_Discussion}

In Figure 10, we compare our resulting visually inspected CoE to the estimates from \citet{Morse1995} and the automated procedure (see Section \ref{sec:AutovVisual} for more details) in Figure \ref{fig:Center_of_Expansion}.
Our estimate is located $9\farcs2$ to the southwest of the oxygen geometric center and $10\farcs8$ to the south and slightly east from the geometric center of the diffuse rim. Notably, these geometric center estimates are $\approx 3.2\sigma$ and $3.7\sigma$ away from our estimate, respectively.  
Our new age estimate of 2770 $\pm$ 500  yr is consistent with the value of $\approx2500$ yr estimated by \citet{Vogt2011} and \citet{Law2020}, as well as the estimate of $2350 \pm 520$ yr made by \citet{Sutherland1995}.



\subsection{Comparison to other O-rich SNRs}
\label{sec:O_SNRs}

With our new proper motion measurements of optically-emitting ejecta of N132D, we expand the number of young ($<3000$ yr) O-rich SNRs with proper motion studies from three to four.\footnote{Work on Puppis A reported by \citet{Winkler10} remains unpublished.} Table \ref{tab:SNR_Prop} shows properties of these remnants (E0102, Cas A, and G292) compared to those of N132D. Figure \ref{fig:SNR_Comp} displays the positions of O-rich knots with respect to their respective CoEs in physical space.

Compared to the three other SNRs, N132D shows the highest degree of spatial asymmetry in the distribution of high velocity ejecta knots. The majority of the knots are seen to the north of the CoE. Given that CSM/ISM in the northwest is associated with higher densities than in the south \citep{Williams2006}, this unique morphology is likely strongly influenced by explosion asymmetry. Further supporting the notion that N132D was an asymmetric explosion in a uniform environment is the ballistic proper motions we measure, the overall blueshift in the Doppler velocities of the ejecta \citep{Lasker1980,Sutherland1995,Morse1995,Vogt2011,Law2020}, asymmetry in the elemental abundances in X-ray observations \citep{Sharda2020}, and evidence of a bipolar explosion from 3D reconstructions \citep{Vogt2011}.

In contrast, the distribution of high velocity ejecta knots in E0102 is fairly uniform, with only the northern part of the remnant lacking any high proper motion knots \citep{Vogt2017}. There is also a notable asymmetry in the proper motions, showing evidence that E0102 is now undergoing non-homologous expansion of optical ejecta  \citep{Banovetz2021}. Proper motion measurements show non-ballisitic motion, with slower material preferentially in the east \citep{Banovetz2021}, suggesting that  E0102 is likely interacting with an inhomogenous surrounding environment. X-ray studies also show varying densities across the remnant, as well as a non-spherical forward shock \citep{Sasaki2006,Xi2019}. While some level of explosion asymmetry may be present, E0102's morphology is most likely dominated by effects from an inhomogenous surrounding environment.

G292, which is elongated along the north-south direction, shows very little evidence of an inhomogenous environment and its shape comes mostly from explosion asymmetry. Both proper motion studies and Doppler measurements show an asymmetric nature to the explosion \citep{Ghavamian2005,Winkler2009}. The proper motion-derived CoE coinciding with the X-ray and radio centers of emission \citep{Winkler2009} also support the notion of minimal CSM interaction. Despite interaction with an equatorial bar of CSM material, overall G292 appears to be expanding into a low density environment \citep{Ghavamian2005}. However, CSM interactions cannot be ruled out. Recent simulations show that inhomogenous surrounding environments are reflected in the forward and reverse shock for only $\approx2000$ yr \citep{Orlando2022}. There is also evidence that G292's morphology is influenced by the motion of its surviving pulsar \citep{Temim2022}.

Cas A shows an asymmetry in the distribution of its highest velocity oxygen knots, although not to the extent of N132D. This remnant shows a main shell of material moving at $\approx$4000 to 6000 \kms that is broadly symmetric in the plane of the sky, but there is also an extended component of sulfur-rich material to the northeast that extends to velocities upwards of $\approx 1.5\times10^{4}$\,km\,s$^{-1}$ \citep{Hammell2008,FM2016}. A complementary high velocity outflow also exists in the southwest \citep{Fesen2001}. Knots of other chemical abundances of Cas A are more symmetrical \citep[see][]{Fesen2006,Hammell2008,MF13}, and although there is a gap of O-rich material in the south of Cas A as seen in Figure \ref{fig:SNR_Comp}, sulfur-rich main shell ejecta at slower velocities is present. Simulations and observations of clumpy, filamentary nebulosities have shown that Cas A likely interacted with an inhomogenous CSM environment \citep{Weil2020,Orlando2022}. Overall, Cas A is a mixture of explosion asymmetry and an inhomogenous surrounding environment.

\section{Conclusion} \label{sec:conclusion}

We present the first proper motion measurements of optical emitting ejecta of SNR N132D in the LMC. The proper motions were measured using manual and automated procedures applied to two epochs of high resolution HST data taken 16 yr apart with the same ACS/F475W instrument+filter combination sensitive to [\ion{O}{3}] $\lambda\lambda$4959, 5007 emission.  With these proper motion measurements, we have increased the number of young, O-rich SNRs with proper motion derived CoEs from three to four.

Our measurement of the CoE made via visual inspection converged on coordinates  $\alpha$=5$^{h}$25$^{m}$01.71$^{s}$ $\delta$=-69$^{\circ}$38$^{\prime}$$41\farcs64$ (J2000) with 1-$\sigma$ uncertainty of $2\farcs90$. Our new CoE estimate is approximately $9\farcs2$ and $10\farcs8$ from previous estimates using geometric centers of emission \citep{Morse1995}. Combining this CoE estimate with the proper motion measurements leads to an age of 2770 $\pm$ 500  yr, consistent with recent age estimates of $\approx2500$ yr by 3D reconstructions of N132D \citep{Vogt2011,Law2020}. 


Our new CoE and explosion age serves as a useful guide for searches to possibly locate the associated neutron star of the original core collapse explosion of N132D \citep[e.g.,][]{Holland2017,Katsuda2018}. To date, no neutron star has been identified in N132D, and our CoE identifies a region where a targeted search can be performed with new 1 Msec Chandra observations (PI: Plucinsky). Our CoE and age estimates of N132D can also effectively guide searches for a surviving binary companion to the progenitor system. To date, there have been no surviving stellar companions found for the population of nearby stripped-envelope SN remnants (see, e.g., \citealt{Kerzendorf2019}). The nearby distance of N132D makes it possible to probe individual stars in the remnant's stellar neighborhood and avoid distance uncertainties and source confusion encountered in studies at extragalactic distances \citep{Fox22}. An attempt was recently made to identify the surviving companion of E0102 \citep{Li2021} using an updated CoE; thus our new CoE for N132D makes the remnant an excellent opportunity for a similar analysis.

\acknowledgements

We thank the anonymous referees and data editor for helping improving this paper. D.~M.\ acknowledges NSF support from grants PHY-1914448, PHY- 2209451, AST-2037297, and AST-2206532. 
C.J.L. acknowledges funding from the National Science Foundation Graduate Research Fellowship under Grant DGE1745303.
This research is based on observations made with the NASA/ESA Hubble Space Telescope obtained from the Space Telescope Science Institute, which is operated by the Association of Universities for Research in Astronomy, Inc., under NASA contract NAS 5-26555. These observations are associated with HST programs 6052, 12001, 12858, and 13378. Support for program \#13378 was provided by NASA through a grant from the Space Telescope Science Institute, which is operated by the Association of Universities for Research in Astronomy, Inc., under NASA contract NAS 5-26555.

\software{PYRAF \citep{PYRAFcite}, ds9 \citep{ds9cite}, astrometry.net \citep{Lang2010}, Astropy \citep{AstropyCiteA,AstropyCiteB}}

\hfill \break

\bibliography{bib}{}
\bibliographystyle{aasjournal}

\appendix
\setcounter{figure}{0}
\renewcommand{\thefigure}{A\arabic{figure}}

\setcounter{table}{0}
\renewcommand{\thetable}{A\arabic{table}}

\section{More Detailed look into the Computer Vision Approach}
\label{sec:Automated}

\subsection{Previous Automation Techniques}
In this paper, we used a new automated procedure for measuring the proper motions of supernova remnant ejecta. Although manual inspection is a reliable method, using computer vision measurement techniques can allow for rapidly reproducible results, testing various quantitative thresholds, and scale to a large number of proper motions more efficiently. One common technique is using a cross-correlation method \citep[e.g.,][]{Currie1996}, that was been used for the proper motion measurements of SNRs \citep[e.g.,][]{Finkelstein2006,Winkler2009}, stellar ejecta \citep[e.g.,][]{Morse2001,Kiminki2016}, and protostellar jets \citep[e.g.,][]{Hartigan2001,Bally2002}. 
This uses predefined box regions to outline a `clump' in one epoch that is then translated to a new position. The translated image is then subtracted by the second epoch and the translation that produces the minimum sum of the differences is the translation that is used.
Other computer vision techniques for measuring proper motions in SNRs include measuring the optical flow, projection methods, and maximum likelihood functions \citep[e.g.,][]{Borkowski2020}. 

The success of these procedures depends on the knots being bright and are best suited for large regions of gas. However, with the high spatial resolution of HST, we are able to resolve individual knots and cover more of the periphery of the SNR where faint, high proper motion ejecta knots are expected to be. Still, these knots can change in illumination and shape between epochs \citep[e.g.][]{Fesen2011,Patnaude2014}, which can lead to errors in tracking.
One of the goals of our new procedure is to be able to track fainter, individual knots for an older O-rich remnant such as N132D.

\subsection{Automated Proper Motion Measurement Procedure}
\subsubsection{Image Preparation}
\label{sec:ImPrep}
The first step in image preparation was to ensure that emission unrelated to N132D's oxygen-rich material was corrected for and removed. Doing so ensured that our computer vision procedure only tracked [\ion{O}{3}] $\lambda\lambda$4959, 5007 emission and was not confused by CSM/ISM or stellar emission that was easily recognized in and avoided by the manually inspected proper motion measurements. 

We first scaled the F550M (continuum) and F658N (H$\alpha$) images taken in 2004 to the ACS/F475W images by matching the flux of common stars and then subtracting the F550M and F658N emission from the ACS/F475W image. This is to remove the stellar continuum and H$\beta$ emission (using H$\alpha$ as a tracer) from our images in order to only track the oxygen-rich ejecta. 
We then manually removed any subtraction residuals outside of emission regions, and performed additional cleaning of the images using \texttt{cosmicray\_lacosmic} task from the \textit{astropy} package \textit{ccdproc}.

\begin{figure*}[tp]
\centering
\includegraphics[width=0.65\linewidth, angle=0]{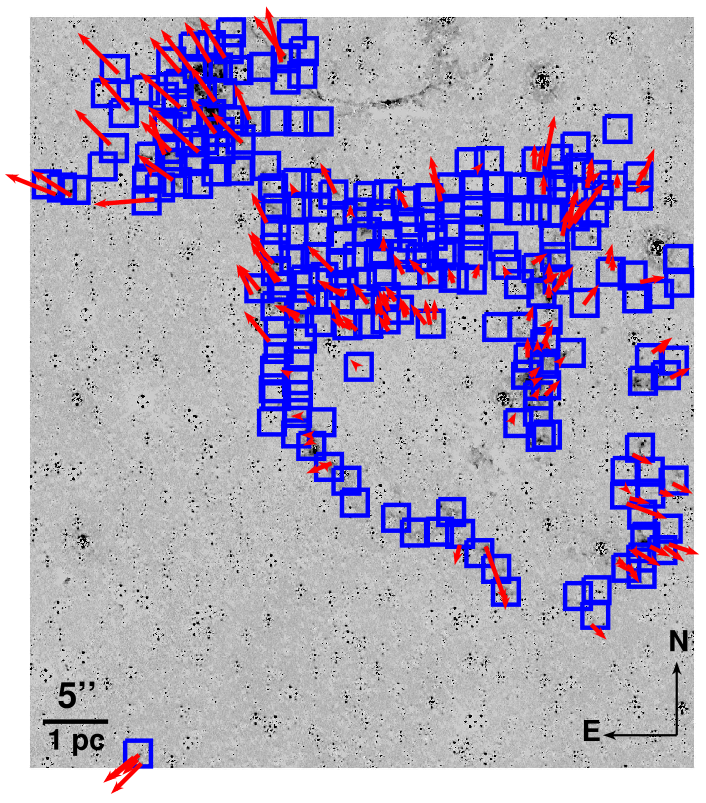}
\caption{2004 ACS/F475W continuum- and hydrogen-subtracted image with vectors representing the automated procedure's measured shifts (multiplied by a factor of 20 for visual clarity) shown in red. The stamps used in the automated procedure are shown as blue boxes. Empty boxes represent areas where the automated procedure could not detect any motion and/or identify knots.
}
\label{fig:Auto_Vector}
\end{figure*}

\subsubsection{Identifying Knots}
\label{sec:IDKnots}
\begin{figure}[tp]
\centering
\includegraphics[width=0.4\textwidth, angle=0]{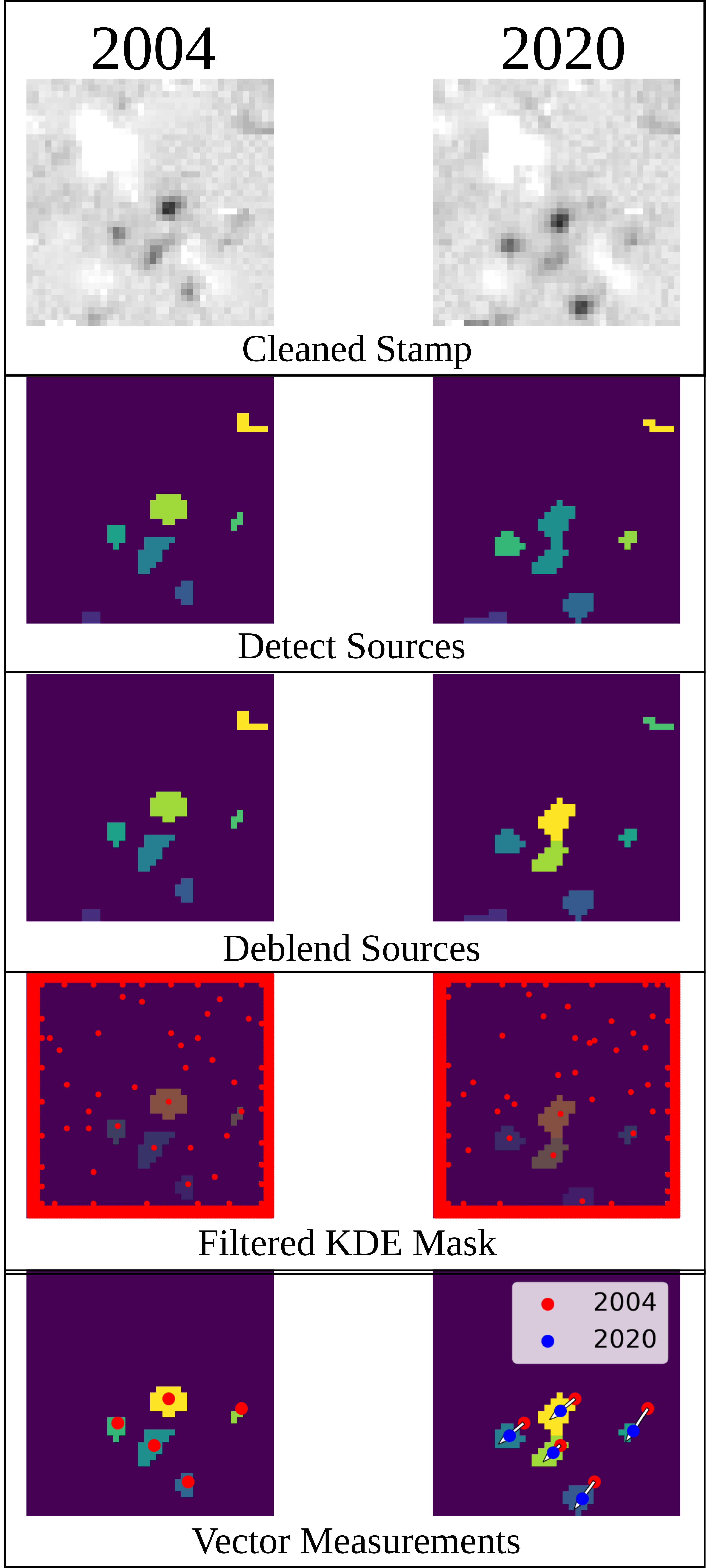}
\caption{Steps needed for the computer vision technique to identify and measure the proper motion of ejecta knots. The 2004 and 2020 images of the same region as Figure \ref{fig:Tracking} are shown on the left and right panels,  respectively. The top row shows the original cutout of the continuum and hydrogen subtracted image in gray scale. The second and third rows show the results of the \texttt{detect\_sources} and \texttt{deblend\_sources} tasks, respectively. 
The fourth row shows the remaining segments after applying the KDE to the total region, identifying the localized peaks as shown in red. The final row shows the matched centroids. The centers are shown in red (2004) and blue (2020), with a white vector showing the motion between the two epochs.}
\label{fig:Auto_procedure}
\end{figure}

Next, small regions were identified for further individual processing. We chose regions that had strong oxygen emission or showed evidence of proper motion measurements when blinking between  epochs. We divide these regions into 2$^{\prime\prime}$ by 2$^{\prime\prime}$ boxes, or stamps. 
Figure \ref{fig:Auto_Vector} shows all of the stamps while the first row of Figure \ref{fig:Auto_procedure} shows an enlarged example of a stamp. The stamp's size was chosen for its ability to encapsulate knot motion between epochs and to sample the local background. This is important since N132D's non-uniform emission properties makes using a global background value impractical. We allowed large overlap between the stamps to ensure that no knots were missed.

We then used the \texttt{detect\_sources} and \texttt{deblend\_sources} tasks from the \textit{photutils} package in \textit{astropy} to identify knots, shown in the second and third row of Figure \ref{fig:Auto_procedure}, respectively. \texttt{Detect\_sources} creates a segmented image which identifies sources of emission above a certain threshold. In our case, this was 2$\sigma$ above the median background of the 2$^{\prime\prime}$ by 2$^{\prime\prime}$ stamp. We further restricted our analysis to those knots containing emission that spanned at least four adjacent pixels. This pixellimit was selected to account for smaller, fainter knots than those found through visual inspection.

\texttt{Deblend\_sources} takes the segmented regions and finds the local maxima to separate potential overlapping knots. For this process, we assumed a Gaussian kernel and a low contrast between the knots. We then apply three initial filters to remove any embedded residuals or hot pixels that were missed with \texttt{cosmicray\_lacosmic}. The first filter removes segments too close to the edges of the region, effectively reducing the region from  2$^{\prime\prime}$ by 2$^{\prime\prime}$ to $1\farcs9$ by $1\farcs9$. The next filter removes segments that do not move more than 1.5 \mas, to remove any remaining residuals. This lower limit was chosen based on the manually inspected lowest proper motions around 3 \mas. The last filter removes any segment that does not have a corresponding segment within $0\farcs3$ around it in the other epoch, to remove residuals or hot pixels that appear in one epoch but not the other.

\subsubsection{Matching Knots and Calculating Shifts}
\label{sec:MatchKnots}

The changing brightness and shape of individual knots over time, in addition to any residual hot pixels, can affect the local peak position and the geometric center determined for each knot. To mitigate these effects, the local peaks in the identified knots are calculated using a KDE fit of the stamp (fourth row of Figure \ref{fig:Auto_procedure}). To achieve sub-pixel accuracy, we found the best results by resampling the stamp from 40x40 pixels to 100x100 before applying the KDE, resulting in an error of 0.4 pixels ($\approx0\farcs2$) for the calculated centers. We also found the best bandwidth for the KDE was the the Silverman kernel \citep{Silverman1982} through manual inspection of the resulting KDE centers.

The next step is to match the knots between the epochs. Each pair of centroids is found using an initial guess of the center of expansion to find the position angle between the knot and the guess. We use the CoE calculated by the visually tracked knots for the initial guess. Each individual pair of knots between epochs is given a score, which is the shift between epochs multiplied by the difference between their trajectory  and the position angle. The pair with the lowest score is selected as the correct combination. The combinations are then passed through two filters to mitigate outliers. Knot combinations are discarded if the difference between the trajectory and the position angle is greater than 90 degrees and if the shift is larger than 30 \mas, double the largest proper motion found using visual measurements. The effect of changing these parameters are explored in Section \ref{sec:AutovVisual}. If multiple combinations use the same endpoint (e.g. the deblending process splits a knot into two between epochs), the combination with the lowest score is chosen and the other is discarded. The final row of Figure \ref{fig:Auto_procedure} shows an example final combinations. 

Finally, we clip the top and bottom $5\%$ of the proper motion vectors to remove outliers, average duplicate measurements due to overlapping stamps, and remove measurements that are within 3 pixels of the mask for subtraction residuals. The remaining measurements are used for the proper motion analysis as was done for the manual inspection in Section \ref{sec:Manual}. We ensured that the procedure is robust using a toy model simulation, detailed in the following section.

\subsubsection{Simulated Proper Motion}
We tested our procedure using a simulated, idealized SNR with ejecta moving ballisitically. We generated two 300 by 300 pixel images at a scale similar to HST with randomized background emission using the \texttt{make\_noise\_image} task in \textit{astropy}'s \textit{photutils} package. The first was an image of 45 knots  randomly placed to concentrically surround a test CoE located at the center of the image. The knots were drawn from a sample of the visually inspected HST knots, the majority of which were in the top third of knot brightness, and were placed using the geometric center of each knot. 
The second image was then created using a different randomized background and placing these knots a certain distance, radially away from the simulated CoE with a proper motion with $v \propto r$. We implemented a scanning procedure to find regions containing the knots. This procedure scanned the image using subdivided regions to 40x40 pixel boxes with overlap between them.\footnote{We did not pass regions through the filters for hot pixels and star removal, as neither of these features were present in the simulated images.}

Our automated procedure recovers all 45 knots. There was an average positional difference of 2.12 pixels ($\approx0\farcs1$ with HST resolution), an average shift difference of 1.53 pixels ($\approx$3.5 \mas), and an average angle difference of 0.12 radians between the inferred and true (simulated) values. We calculated a CoE of (154.77,155.97) with a 1$\sigma$ error radius of 10.88 pixels ($\approx0\farcs58$) as compared to the simulated CoE of (150,150).
Overall, we were able to find all of the knots, match them correctly, and calculate their speed and trajectory to return a CoE that is within 1-$\sigma$ of the true value.

\begin{figure*}[tp]
\centering
\includegraphics[width=0.7\linewidth, angle=0]{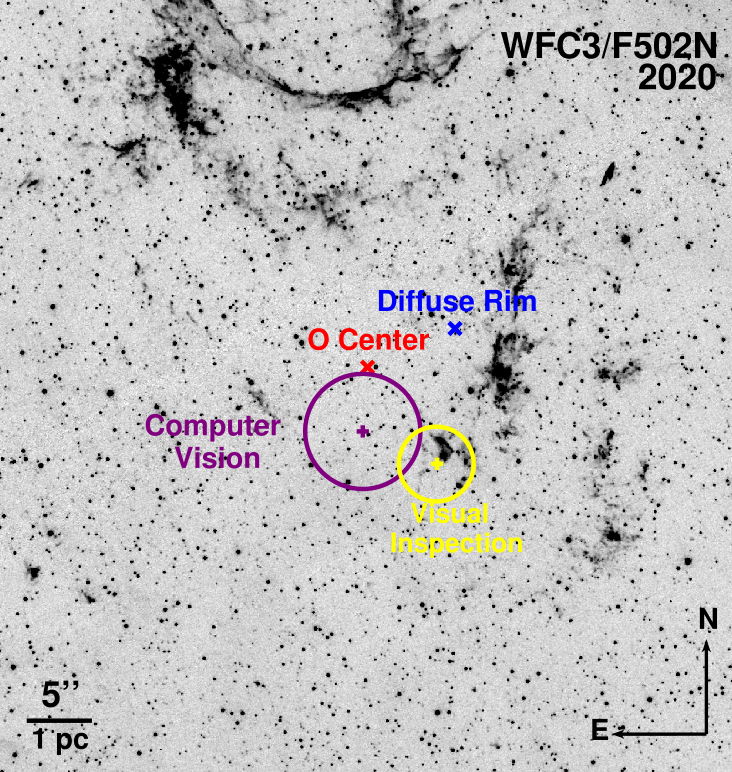}
\caption{Similar to Figure \ref{fig:Center_of_Expansion} with the addition of the automated procedure's result of $\alpha$=5$^{h}$25$^{m}$02.771$^{s}$
and $\delta$=-69$^{\circ}$38$^{\prime}$$38\farcs985$ (J2000) with 1-$\sigma$ uncertainty of $4\farcs47$ in purple.}
\label{fig:Center_of_Expansion_CV}
\end{figure*}

\subsection{Automated Procedure Results}

This procedure was able to identify and track 137 knots of ejecta with the error of the proper motions was set to 0.4 pixels ($\approx0\farcs2$), from the sub-pixel ratio of the KDE. Figures \ref{fig:Auto_Radial_Knot}--\ref{fig:Variation_Age} show the knot locations, trajectory, and proper motion trends using this procedure. The procedure measured proper motions ranging from 2 to 17 \mas and an $S$ of $0\farcs015$ per \kms.

Using the same CoE method as outlined in Section \ref{sec:CoE}, the 137 proper motions yields a CoE of 
$\alpha$=5$^{h}$25$^{m}$02.771$^{s}$
and $\delta$=-69$^{\circ}$38$^{\prime}$$38\farcs985$ (J2000) with 1-$\sigma$ uncertainty of $4\farcs47$. Figure \ref{fig:Center_of_Expansion_CV} in the Appendix shows this result as compared to the other center of expansion estimates.

Utilizing this CoE and proper motions, we calculate an age of 3377 $\pm$ 2241 yr using all 137 knots, as shown in the left panel of Figure \ref{fig:Auto_years}. This large discrepancy is most likely due to the procedure measuring the proper motions of artifacts or heavily decelerated knots, skewing the age estimates to higher values. To account for the decelerated knots, we also calculated the age using knots above the median proper motion. These 70 knots yield an age of 2497 $\pm$ 638 yr (right panel of Figure \ref{fig:Auto_years}), much closer to that derived from visually measured proper motions.

\subsection{Effect of Tuning Parameters}

Figures \ref{fig:Auto_Radial_Knot}, \ref{fig:Center_of_Expansion_CV}, and \ref{fig:Auto_years} show the results of using the conservative metrics to measure the proper motions, their trajectories, and subsequent CoE and explosion ages, respectively. These results are highlighted in Section \ref{sec:AutovVisual}. 
The conservative parameter constraints used in the selection of knots adopted in our automated procedure were used to incorporate many degrees of freedom. However, the
associated proper motion measurement uncertainties were much larger than those associated with our manual procedure. There are many ways that the parameters of knot selection can be further constrained to reduce the uncertainties and better match the manually measured CoE and age. We explored limiting the difference angle between the trajectory and position angle of the input CoE to 45 degrees (from 90), increasing the minimum proper motion to 3 \mas (from 1.5 \mas), and using brighter knots by increasing the signal to noise of selected knots from 2$\sigma$ to 3$\sigma$. We found that by tightening the boundaries of these parameters, the CoE of the automated procedure and the age calculation were both within 1-$\sigma$ of the visual inspection results. The following section contains a detailed discussion about the effect of each of these parameters.

This procedure is ideal for ejecta proper motion analysis of other SNRs. Two parameters that must be changed between SNRs are the KDE bandwidth and the arbitrary CoE.
The KDE bandwidth is very sensitive and needs to be fine-tuned depending on the SNR. A bandwidth that is too small will identify many flux peaks within a knot while a bandwidth that is too large can miss fainter knots.

Another parameter that we explored was the influence of the choice of the arbitrary CoE for the trajectory versus position angle cutoff. For our procedure outlined in Section \ref{sec:MatchKnots}, we chose the CoE calculated using the visual inspection method. Assuming this CoE was not available, we could have chosen the [\ion{O}{3}] geometric center for our arbitrary CoE \citep{Morse1995}. We experimented with the 45 and 90 degree cutoff with this arbitrary CoE and found that the resulting CoE did not match that found with the visually inspected CoE using an arbitrary initial CoE. However, we found that by iterating the arbitrary initial CoE calculation procedure, the CoE calculation converges to the same CoE as if using the visually inspected CoE for the arbitrary CoE. Running through 10 initial guess estimates of the CoE, we found that using a 45 degree cutoff would take 3--4 iterations, whereas the 90 degree cutoff would take 2 iterations before converging on the the same CoE as found by using the visual inspection CoE as the arbitrary CoE. 
As such, we would recommend anyone using our procedure to use the iteration method to verify results, especially if using a geometric center for the arbitrary CoE, as they are often offset from proper motion derived centers \citep[see][]{Thorstensen2001,Katsuda2018,Banovetz2021}.

\subsubsection{Further Fine-tuning of Parameters}

\begin{figure*}[!htp]
\centering
\includegraphics[width=0.95\textwidth, angle=0]{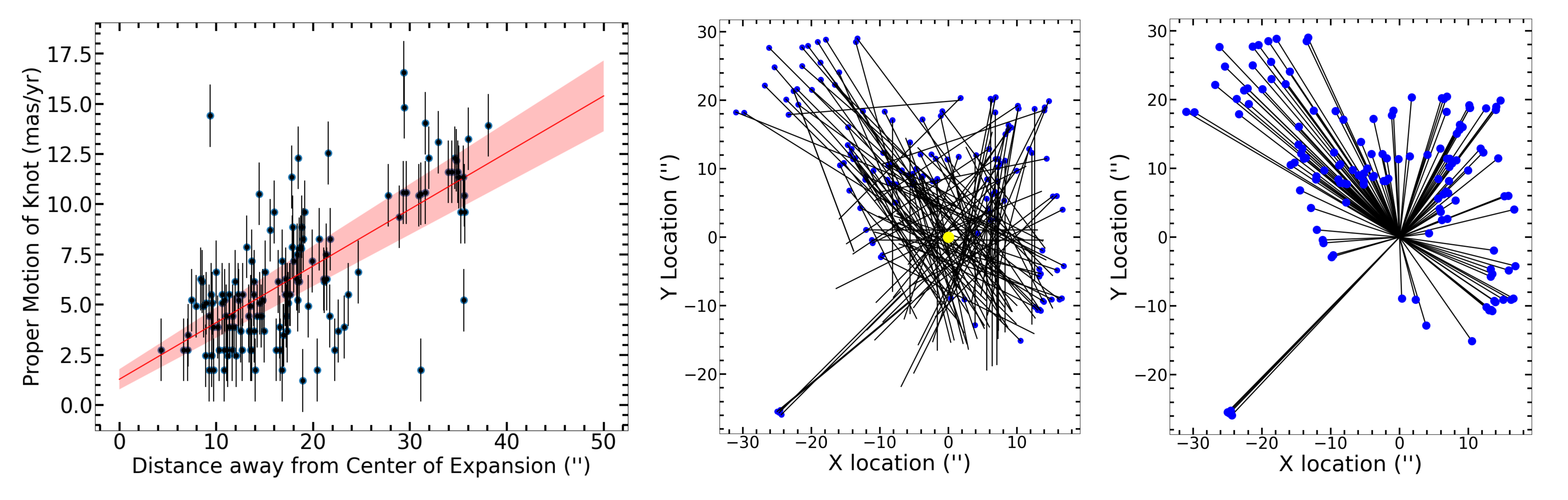}

\caption{Similar to Figure \ref{fig:V_Radial} (left) and \ref{fig:Fesen} (middle and right) but using the proper motions from the conservative parameters of the automated procedure. }
\label{fig:Auto_Radial_Knot}
\end{figure*}

\begin{figure*}[!htp]
\centering
\includegraphics[width=0.47\textwidth, angle=0]{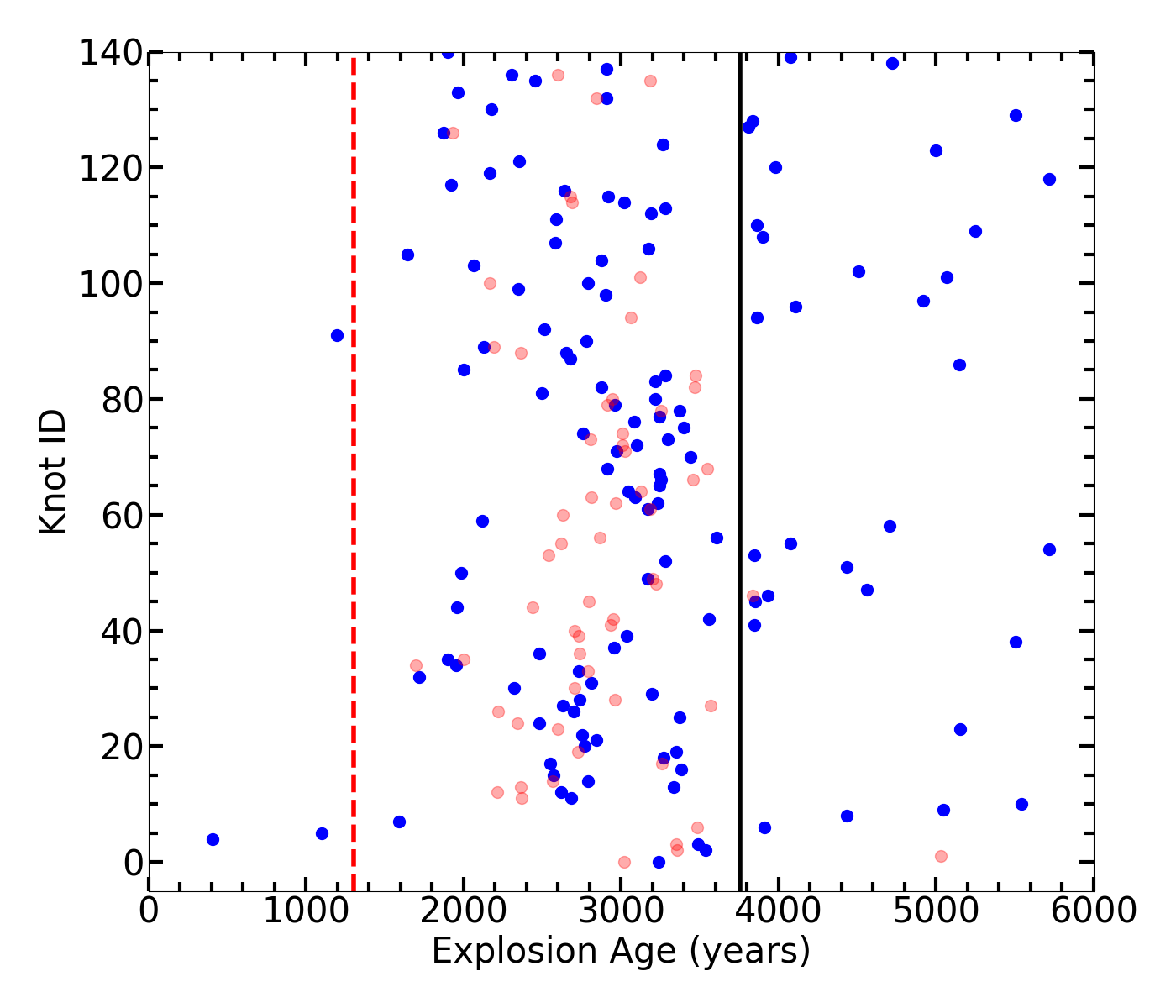}
\includegraphics[width=0.47\textwidth, angle=0]{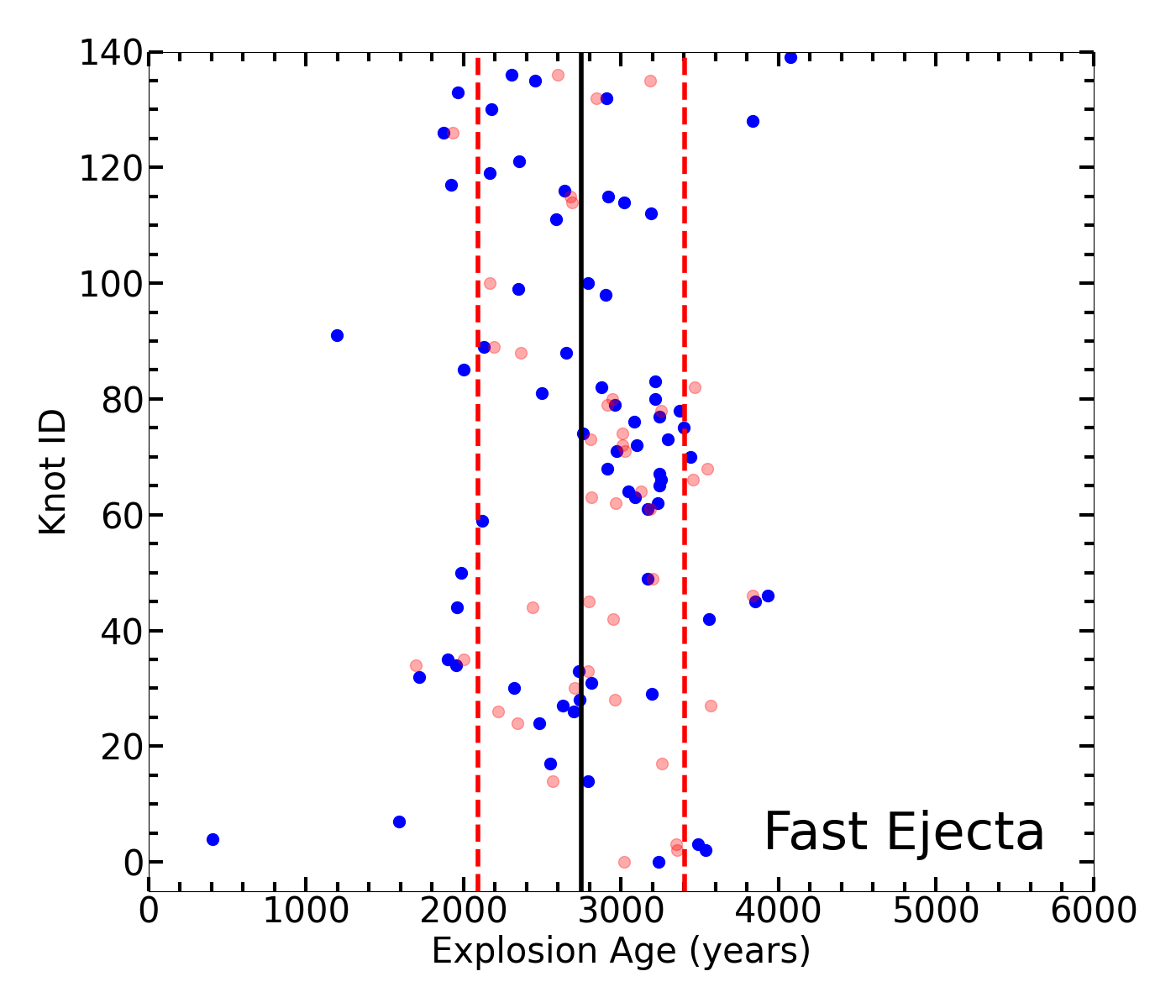}
\caption{Similar to Figure \ref{fig:Fesenyears} but using the proper motions from the automated procedure and including the results of only using the fastest ejecta. This results in an age of 3377 $\pm$ 2241 yr using all the knots and 2497 $\pm$ 638 yr using the fastest. The red shaded points correspond to the matching visually measured knots when applicable.}
\label{fig:Auto_years}
\end{figure*}

Here we discuss how measurement and calculation of the CoE and age are affected by knot selection parameters in our automated procedure. Figure \ref{fig:Auto_Radial_Knot} shows the results of using the conservative parameters for the automated procedure. Figure \ref{fig:Auto_years} show the results of the age calculation when using all 137 knots, or only the fastest (70).

\begin{figure*}[!thb]
\centering
\includegraphics[width=0.24\textwidth, angle=0]{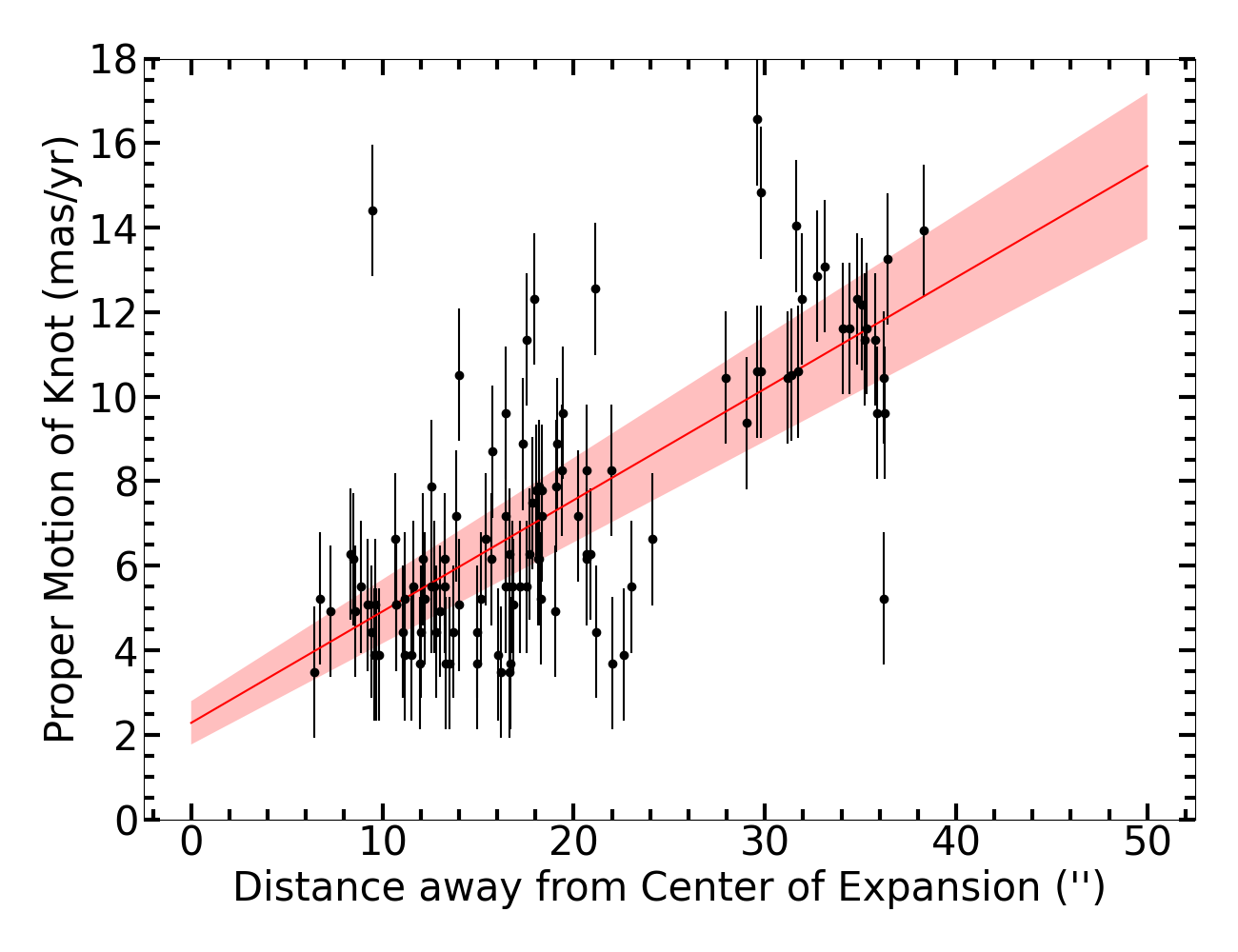}
\includegraphics[width=0.24\textwidth, angle=0]{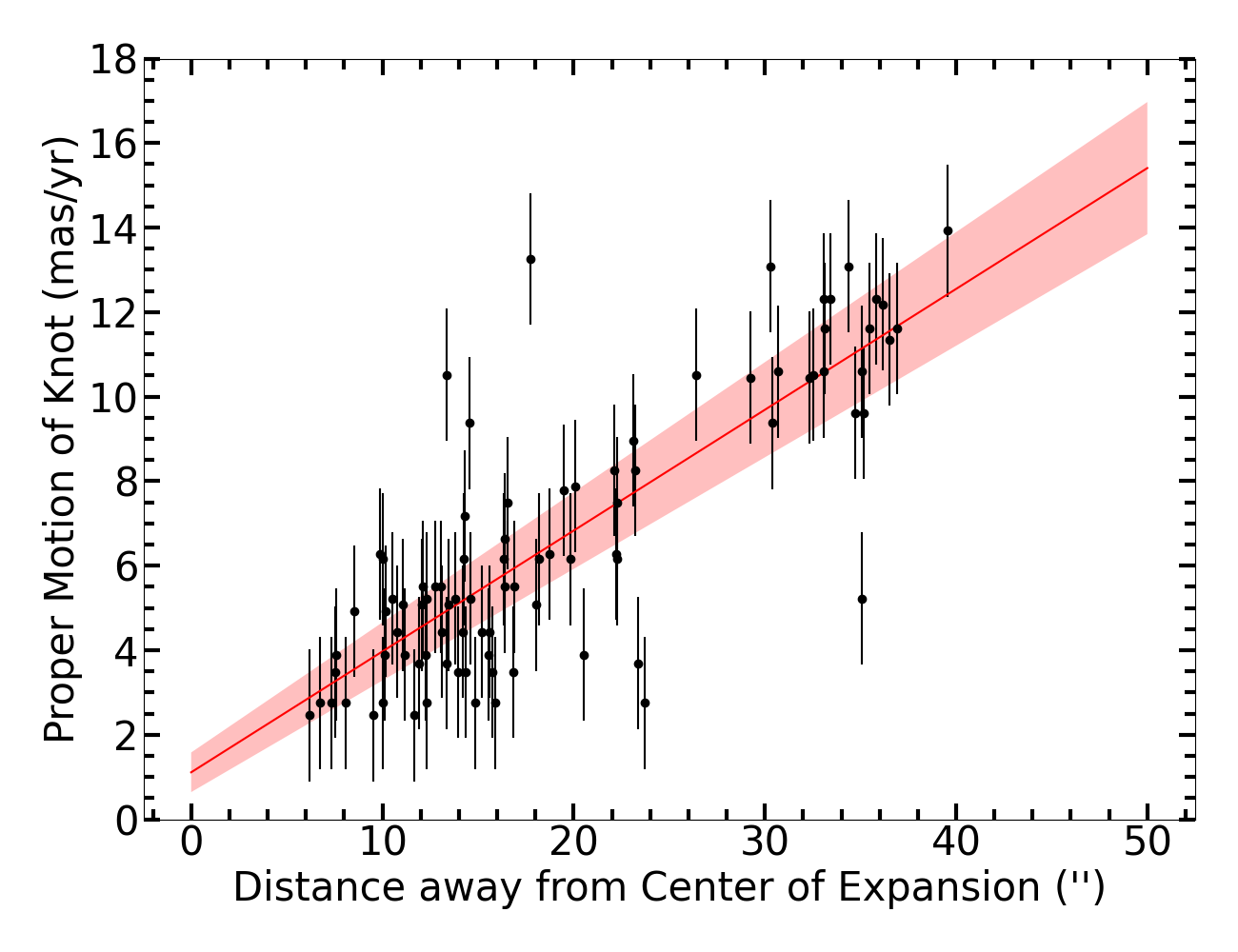}
\includegraphics[width=0.24\textwidth, angle=0]{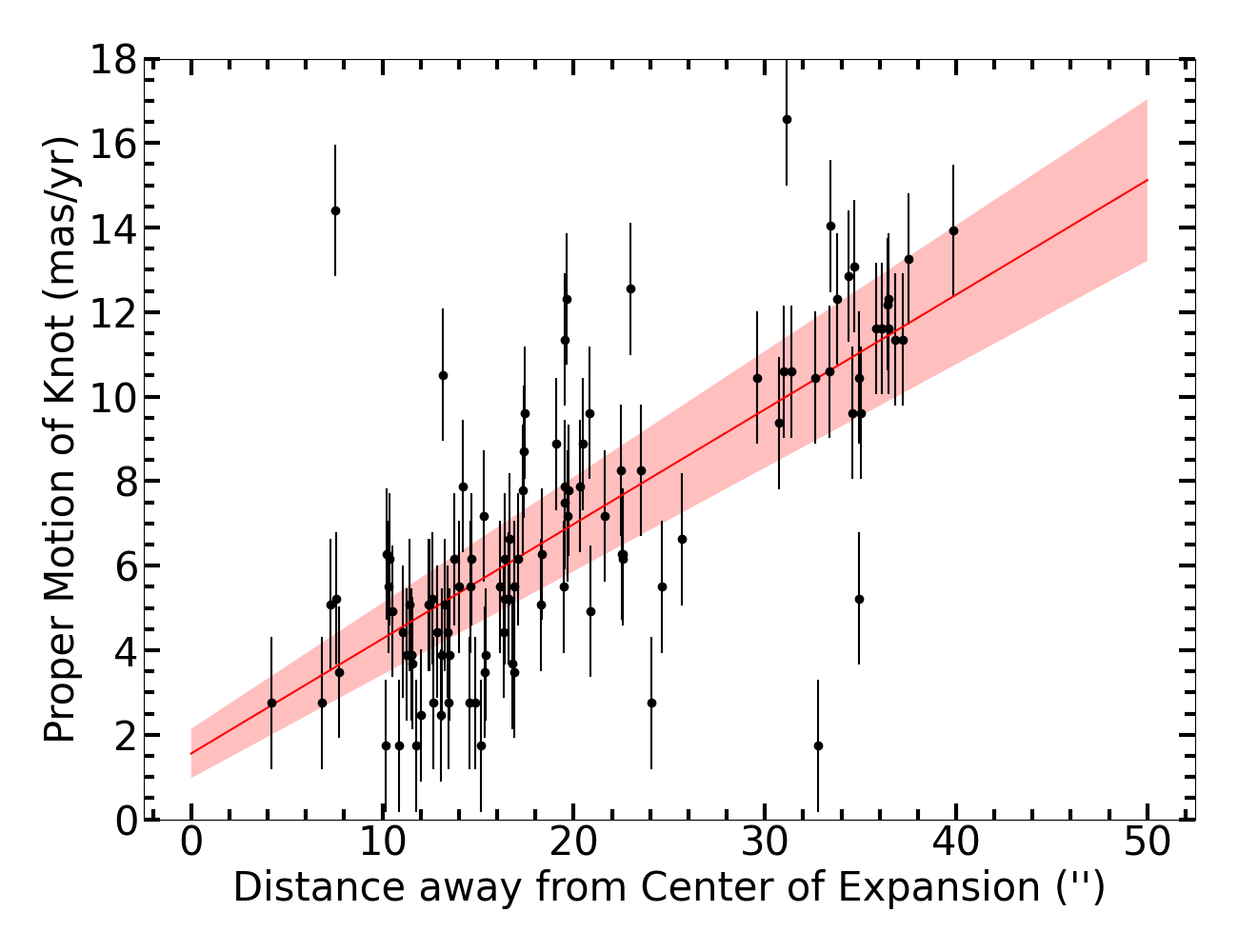}
\includegraphics[width=0.24\textwidth, angle=0]{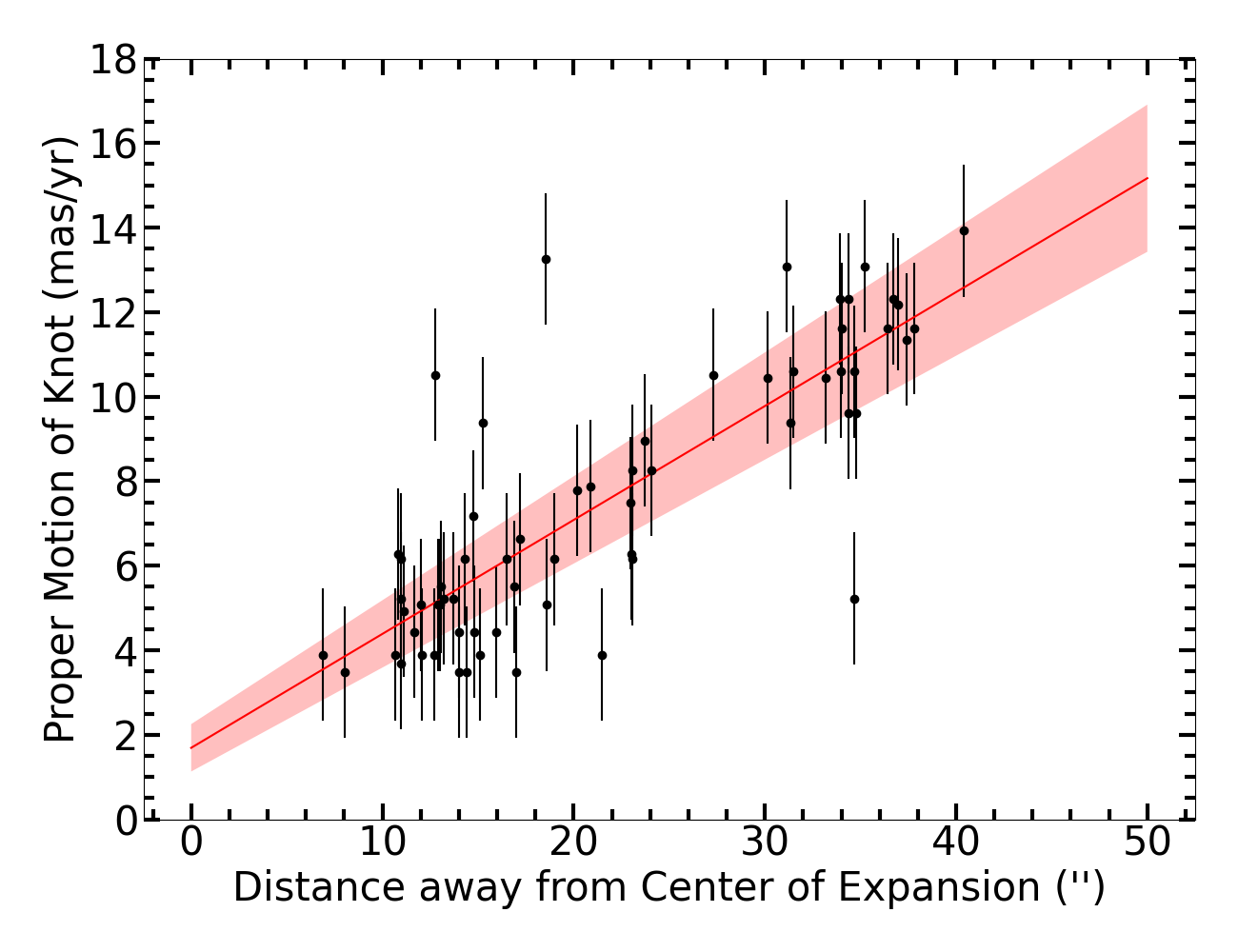} \\

\includegraphics[width=0.24\textwidth, angle=0]{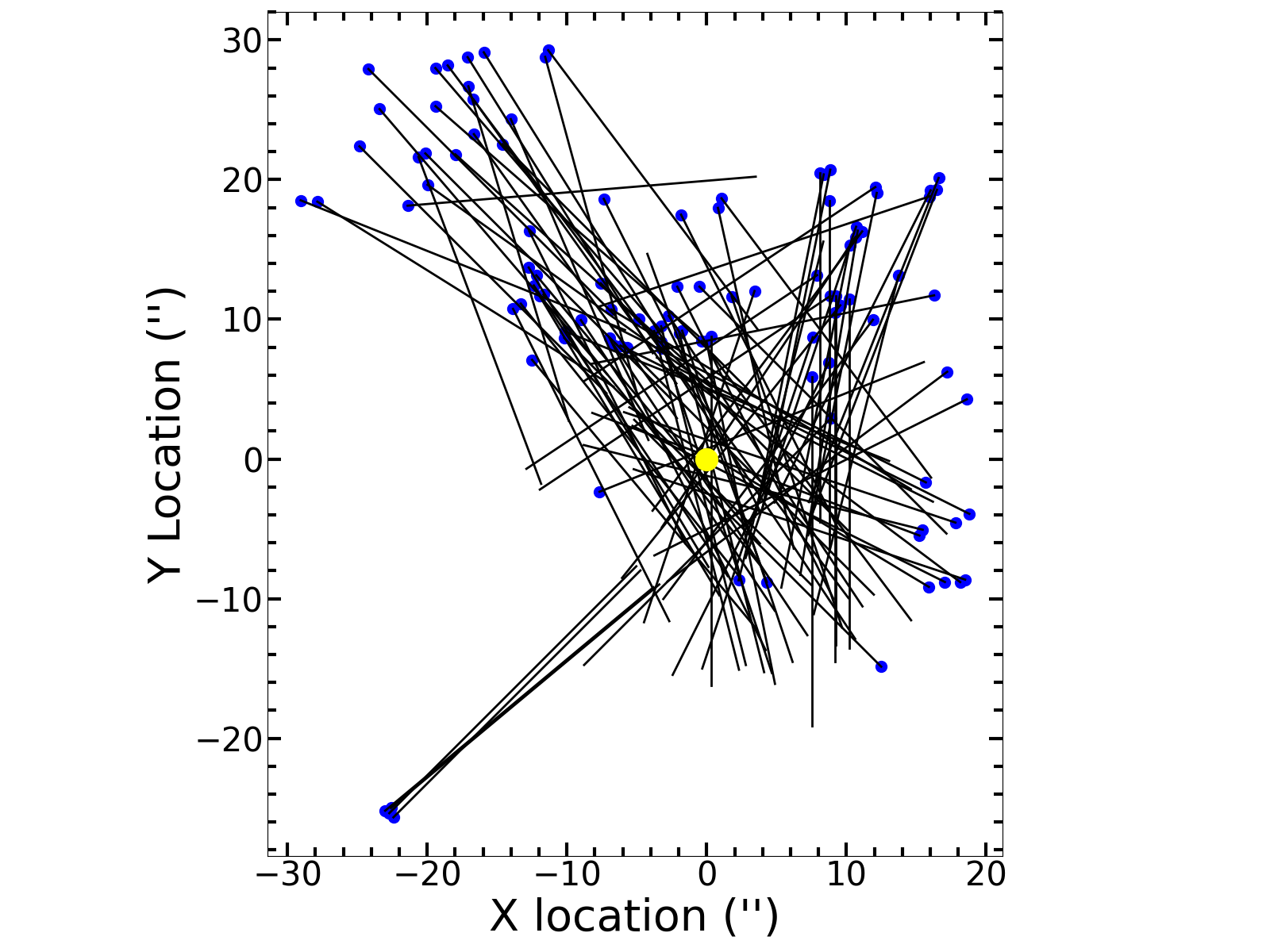}
\includegraphics[width=0.24\textwidth, angle=0]{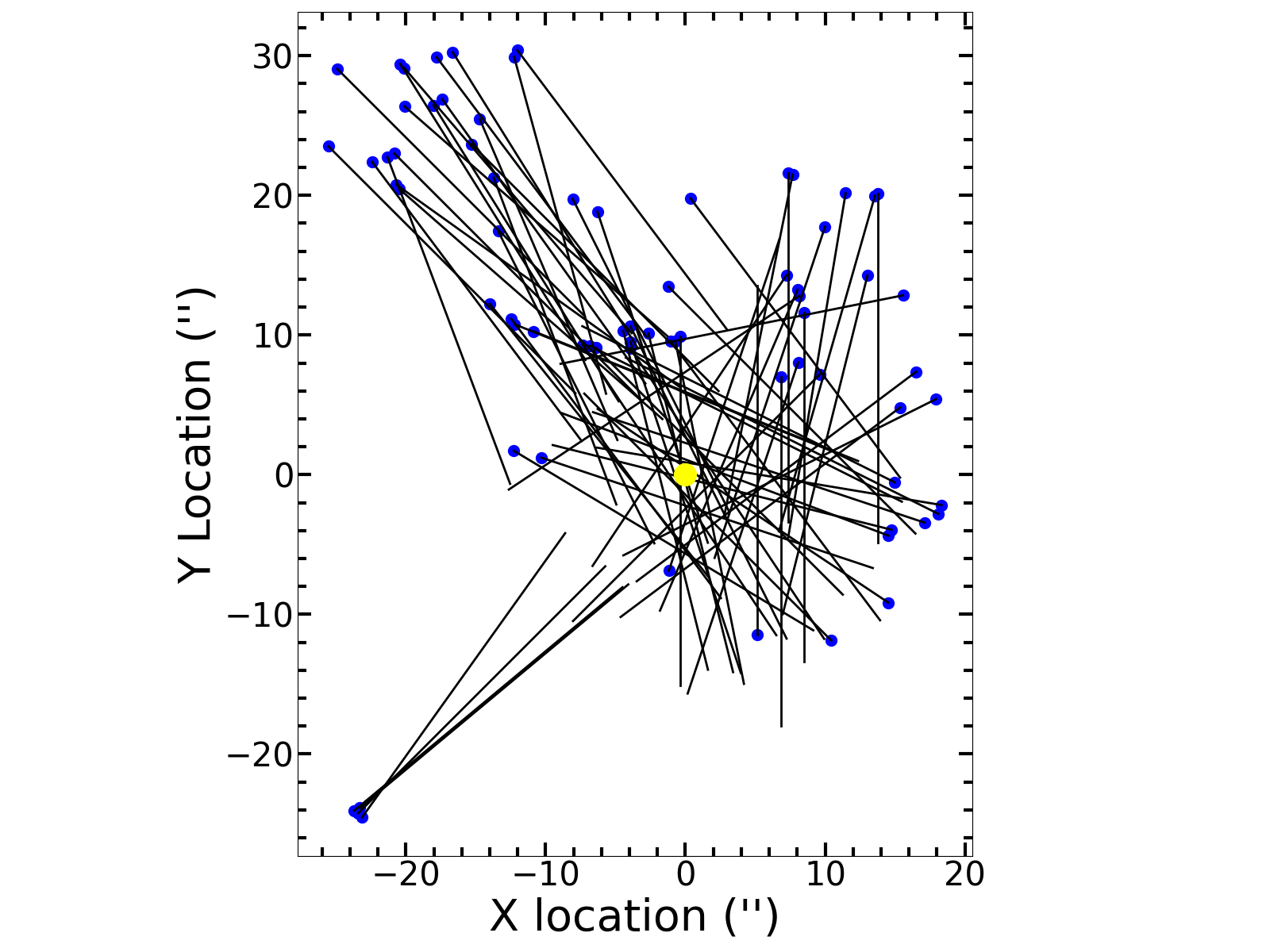}
\includegraphics[width=0.24\textwidth, angle=0]{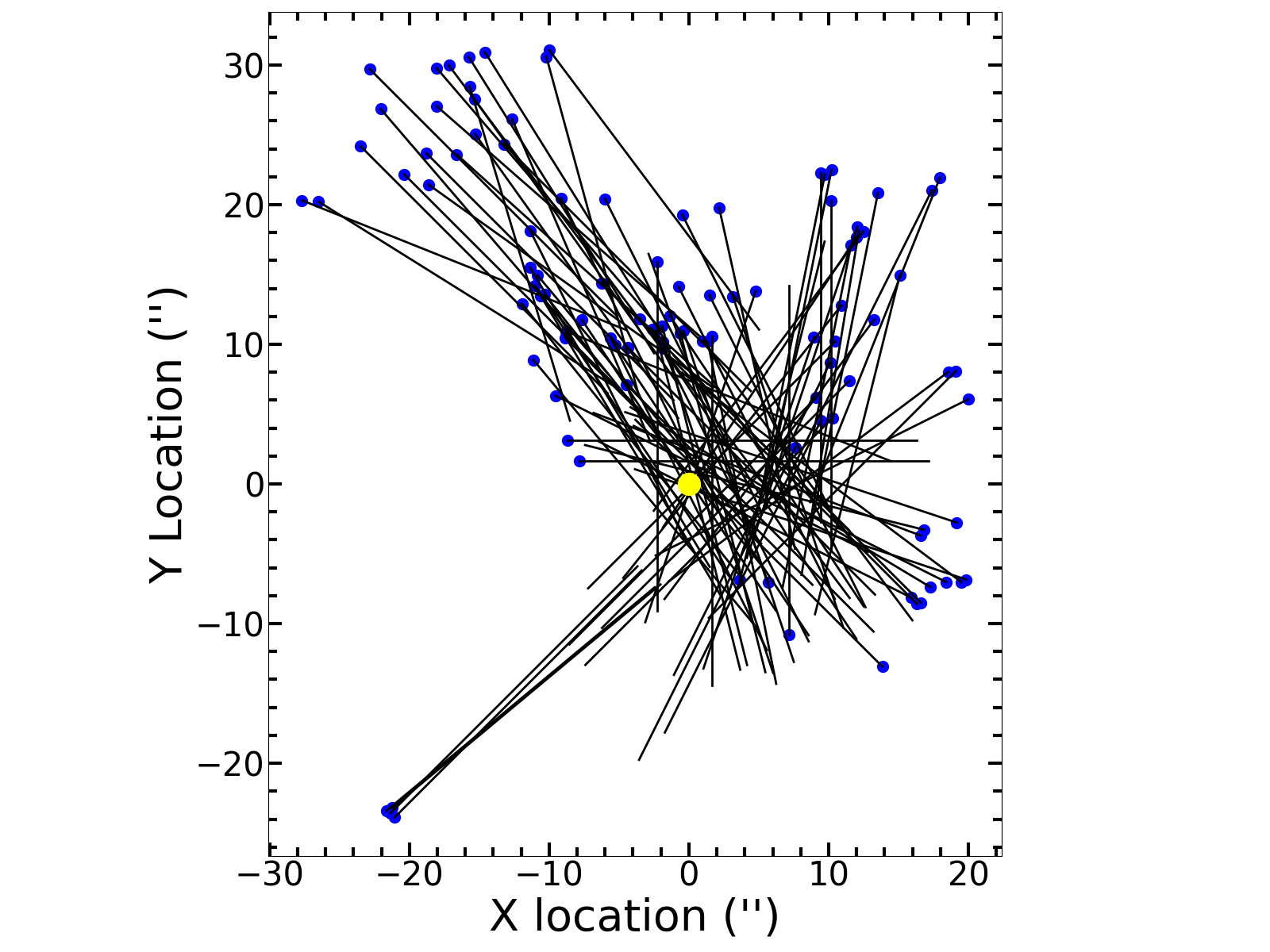}
\includegraphics[width=0.24\textwidth, angle=0]{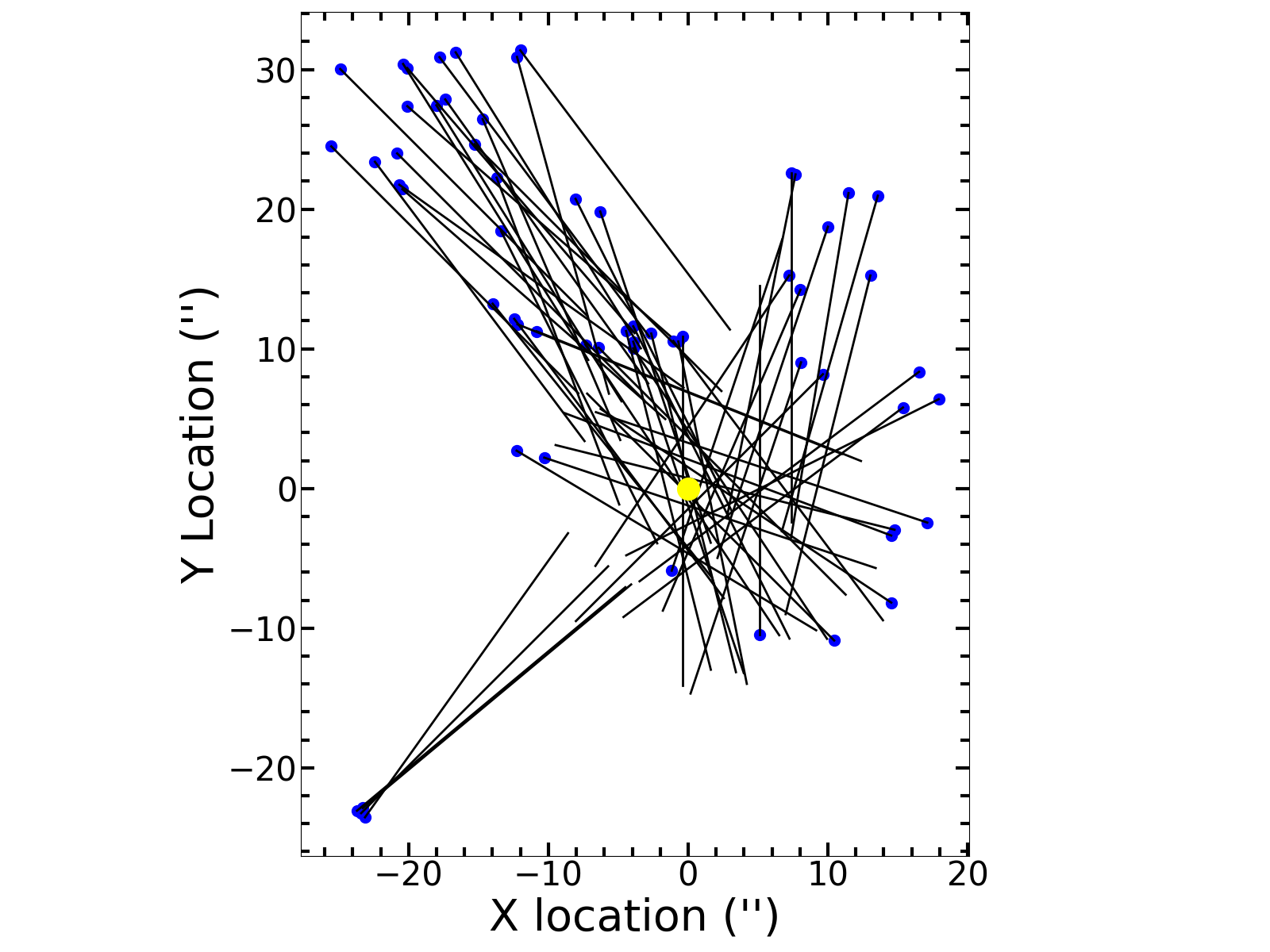}

\caption{Similar to Figure \ref{fig:Auto_Radial_Knot} for the 3 \mas lower limit (left), increase to 3-$\sigma$ of brightness (middle left), and reducing the trajectory and position angle difference of the knots to less than 45 degrees (middle right), and applying all three variations (right).}
\label{fig:Variation_Radial}
\end{figure*}

\begin{figure*}[!t]
\centering
\includegraphics[width=0.40\textwidth, angle=0]{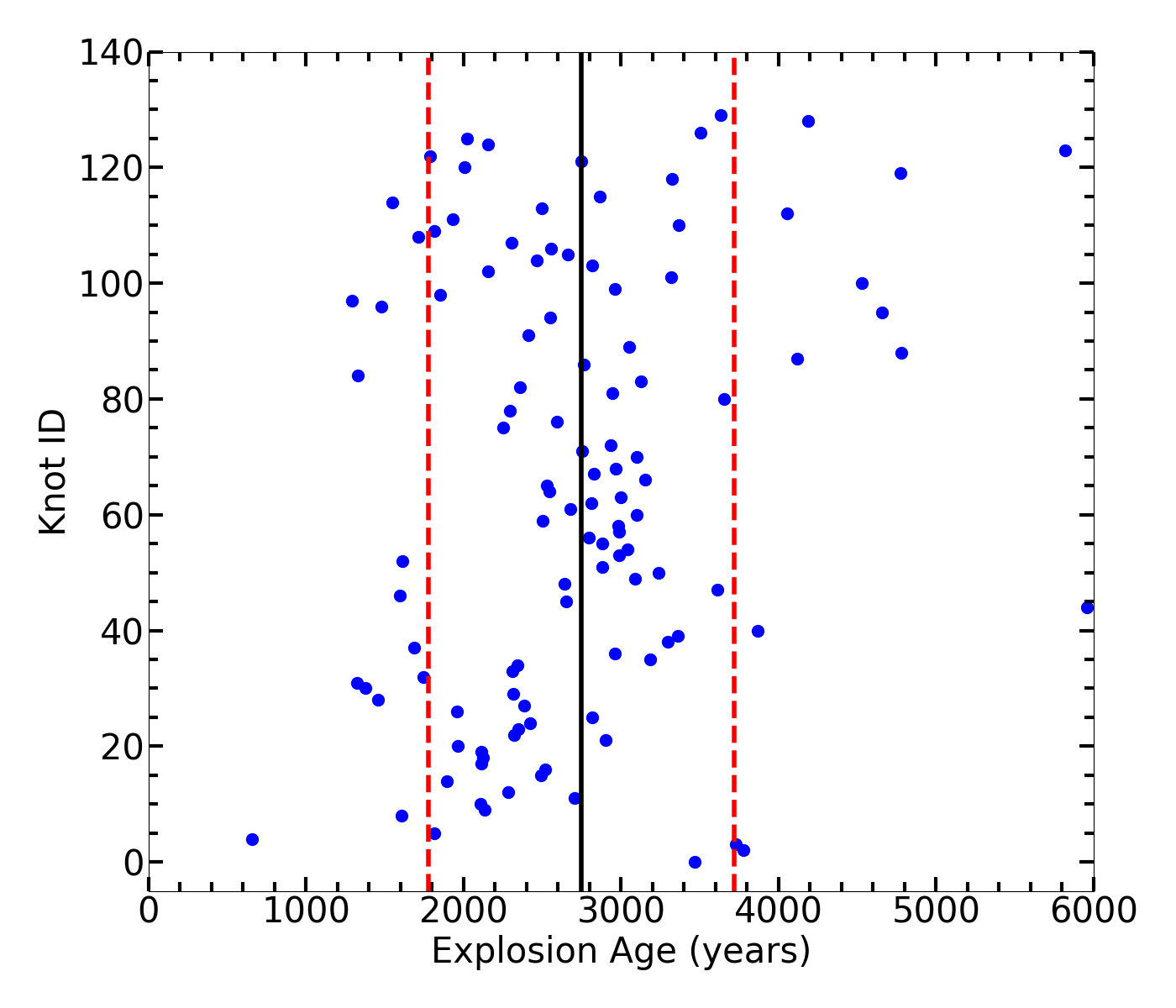}
\includegraphics[width=0.40\textwidth, angle=0]{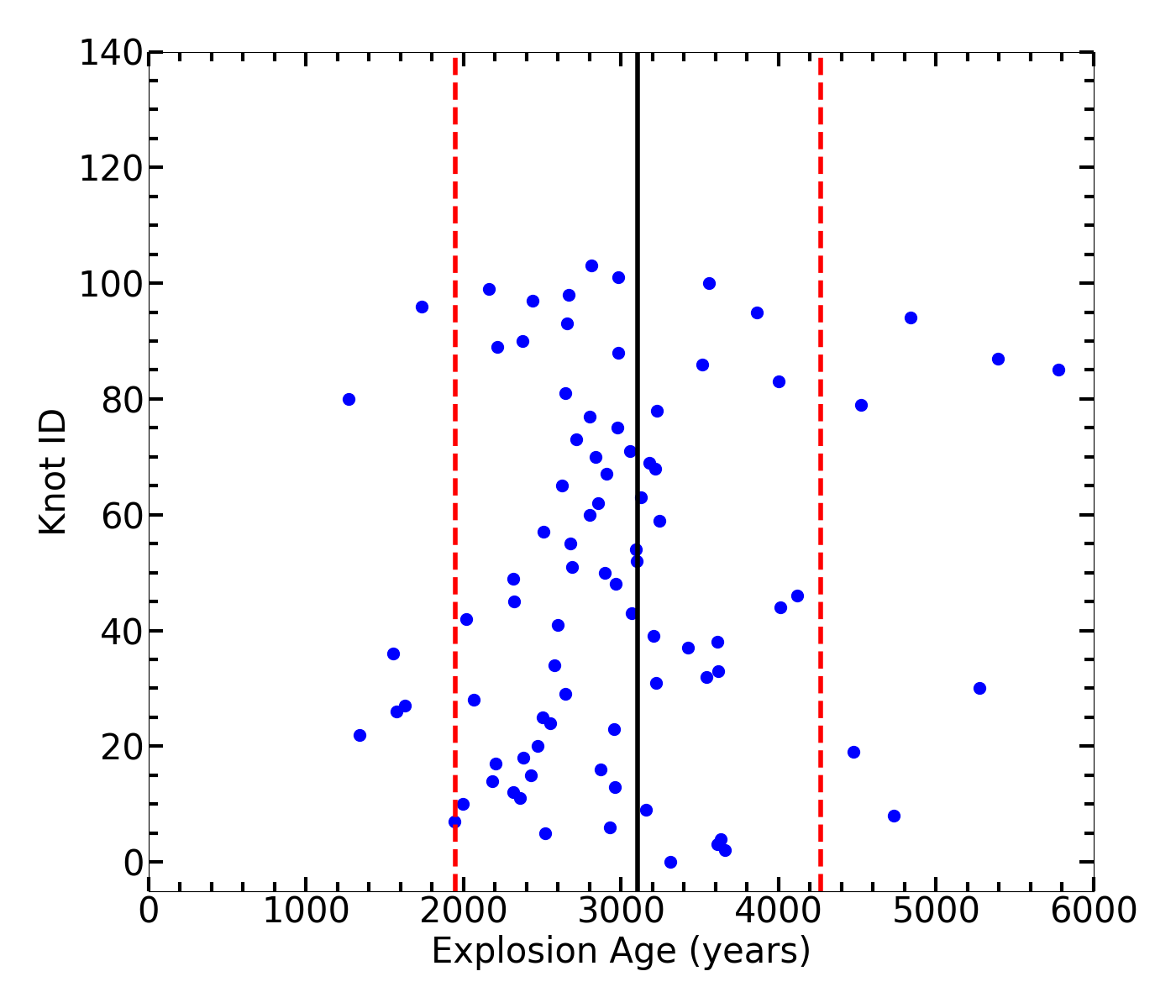} \\
\includegraphics[width=0.40\textwidth, angle=0]{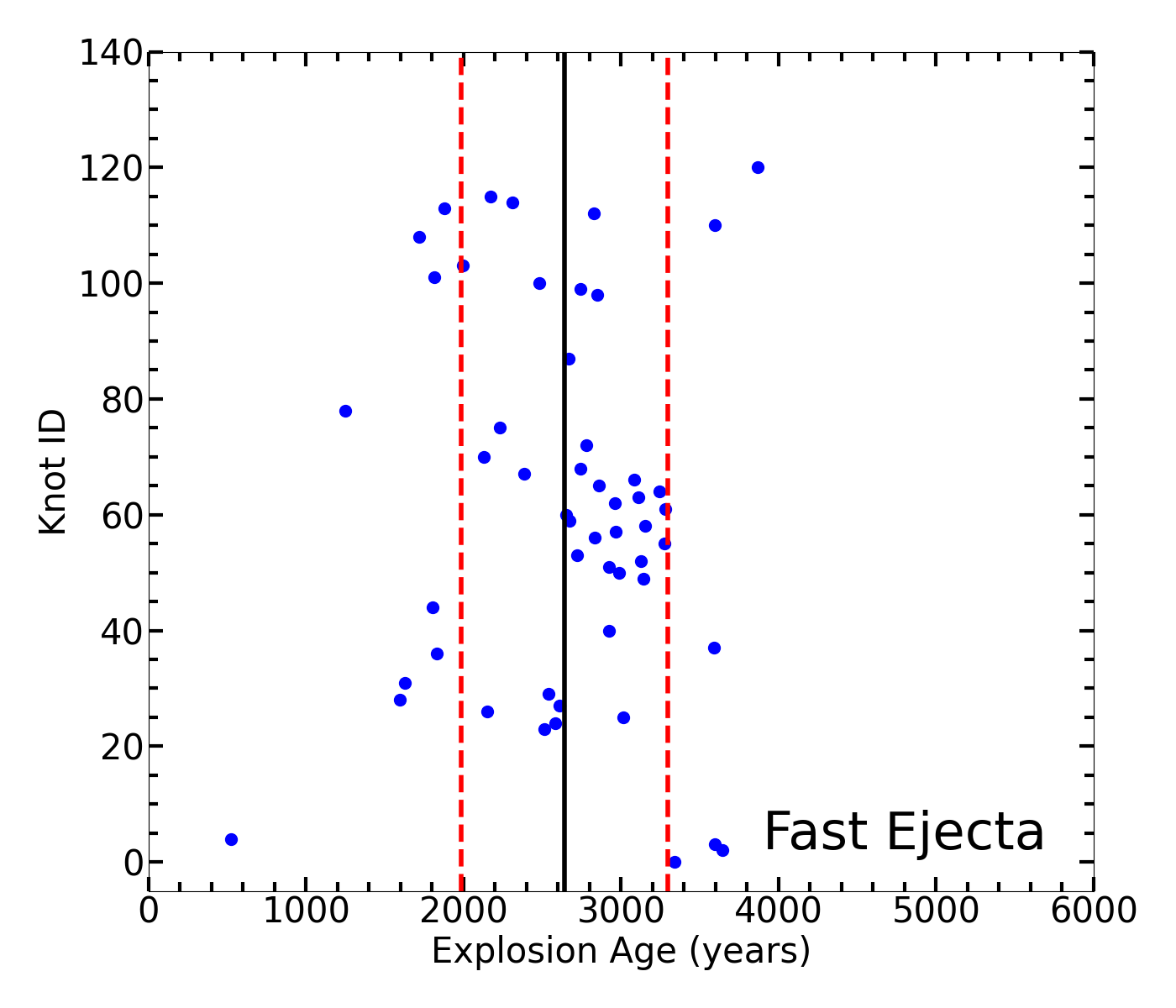} 
\includegraphics[width=0.40\textwidth, angle=0]{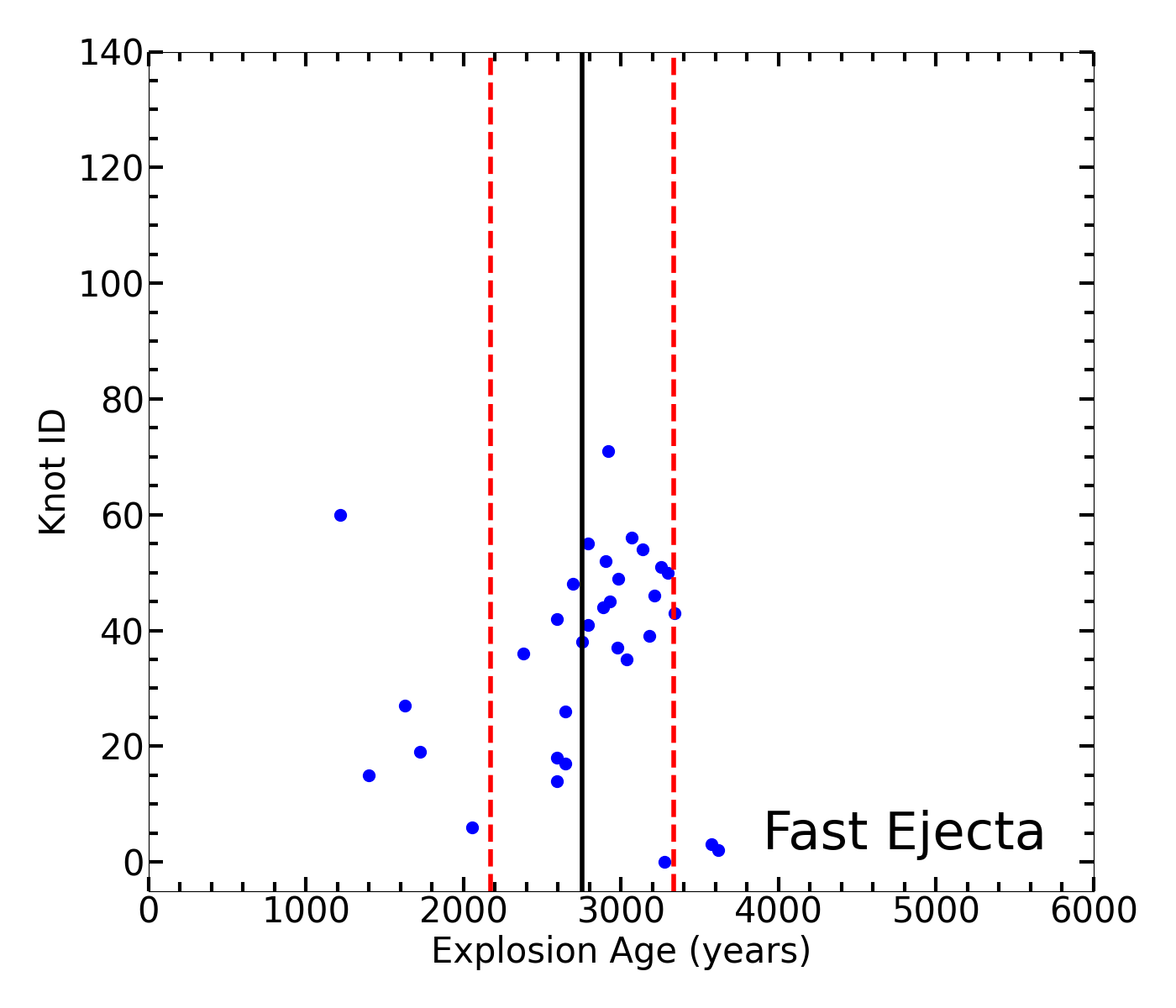} \\
\caption{Similar to Figure \ref{fig:Fesenyears} for the 3 \mas lower limit (top left), increase to 3-$\sigma$ of brightness (top right),the trajectory and position angle difference of the knots to less than 45 degrees (bottom left), and using all the parameters (bottom right) that match closest to the visual inspection result.}
\label{fig:Variation_Age}
\end{figure*}

Figures \ref{fig:Variation_Radial} and \ref{fig:Variation_Age} show the results of changing the parameters of the automated procedure.  The left, middle left, middle right, and right panels show the effect of changing the parameters to incorporate a higher minimum speed (3 \mas), only bright knots (3$\sigma$), difference of trajectory and position angle of 45 degrees, and all three of the changes, respectively.

\clearpage
\section{Additional Tables}
\begin{deluxetable*}{c|cc}[!hp]
\label{tab:stars}
\tablecaption{Anchor Star Coordinates}
\tablehead{
\colhead{Star} & \colhead{RA (J2000)} & \colhead{Dec. (J2000)}
}
\startdata
1 & 5h24m53.7308s & 69d38m15.310s \\ 
2 & 5h24m58.3686s & 69d38m35.298s \\ 
3 & 5h24m57.8822s & 69d38m46.447s \\ 
4 & 5h24m56.5386s & 69d39m38.340s \\ 
5 & 5h24m58.2838s & 69d39m39.118s \\ 
6 & 5h24m42.9113s & 69d39m13.996s \\ 
7 & 5h24m47.0373s & 69d38m43.292s \\ 
8 & 5h24m45.0608s & 69d38m41.959s \\ 
9 & 5h24m49.4502s & 69d38m29.224s \\ 
10 & 5h25m04.9414s & 69d39m39.441s \\ 
11 & 5h25m00.6230s & 69d38m00.893s \\ 
12 & 5h24m51.6282s & 69d39m05.757s \\ 
13 & 5h24m49.5039s & 69d38m47.073s \\ 
14 & 5h24m50.4438s & 69d39m48.192s \\ 
15 & 5h24m52.2404s & 69d39m22.696s \\ 
16 & 5h25m01.3169s & 69d39m48.778s \\ 
17 & 5h24m44.4366s & 69d39m09.882s \\ 
18 & 5h24m56.5344s & 69d38m47.148s \\ 
19 & 5h25m04.2141s & 69d40m09.653s \\ 
20 & 5h24m58.9991s & 69d40m16.083s \\ 
21 & 5h25m07.3918s & 69d40m02.308s \\ 
22 & 5h25m06.7103s & 69d39m40.337s \\ 
23 & 5h25m11.2145s & 69d38m20.227s \\ 
24 & 5h24m45.6713s & 69d38m35.164s \\ 
25 & 5h24m47.2420s & 69d38m17.491s \\ 
26 & 5h24m59.4994s & 69d40m08.583s \\ 
27 & 5h25m08.2993s & 69d39m18.289s \\ 
28 & 5h24m59.8352s & 69d40m17.268s \\ 
29 & 5h24m49.7550s & 69d38m06.447s \\ 
30 & 5h24m52.6476s & 69d38m01.440s \\ 
\enddata
\end{deluxetable*}

\startlongtable
\begin{deluxetable*}{lcccccccc}
\tablecaption{Manually Inspected Knot Measurements and Corresponding Automated Measurements \label{tab:knots}}
\tablehead{
\colhead{Knot} & \colhead{RA} & \colhead{Dec} & & \colhead{Visual} & \colhead{Procedure} & & \colhead{Automated} & \colhead{Procedure}
\\
& & & \colhead{$\mu_{\alpha}$} & \colhead{$\sigma_{\mu_{\alpha}}$} & \colhead{$\mu_{\delta}$} & \colhead{$\sigma_{\mu_{\delta}}$} & \colhead{$\mu_{\alpha}$}  & \colhead{$\mu_{\delta}$} \\
& \colhead{(J2000)} & \colhead{(J2000)} & \colhead{(mas\,yr$^{-1}$)} & \colhead{(mas\,yr$^{-1}$)} & \colhead{(mas\,yr$^{-1}$)} & \colhead{(mas\,yr$^{-1}$)} & \colhead{(mas\,yr$^{-1}$)} & \colhead{(mas\,yr$^{-1}$)}
}
\startdata
1 & 5h25m06.759s & 69d39m04.815s & -8.095 & 0.452 & -7.722 & 0.431 & -7.386  & -7.386  \\ 
2 & 5h25m06.817s & 69d39m04.523s & -5.075 & 0.472 & -4.399 & 0.409 & -3.693  & -3.693  \\ 
3 & 5h25m06.791s & 69d39m04.138s & -7.498 & 0.469 & -6.602 & 0.413 & -7.386  & -6.155  \\ 
4 & 5h25m06.870s & 69d39m04.341s & -7.866 & 0.487 & -6.336 & 0.392 & -7.386  & -6.155  \\ 
5 & 5h24m58.937s & 69d38m48.660s & 5.251 & 0.571 & -2.335 & 0.254 & NA & NA  \\ 
6 & 5h24m59.498s & 69d38m49.597s & 2.766 & 0.412 & -3.161 & 0.470 & 2.462 & -3.693  \\ 
7 & 5h24m59.334s & 69d38m50.789s & 3.786 & 0.403 & -4.496 & 0.478 & NA & NA \\ 
8 & 5h24m59.263s & 69d38m49.396s & 5.096 & 0.520 & -3.409 & 0.347 & NA & NA \\ 
9 & 5h24m59.365s & 69d38m49.148s & 5.557 & 0.543 & -3.162 & 0.309 & NA & NA \\ 
10 & 5h24m59.317s & 69d38m49.096s & 6.207 & 0.550 & -3.349 & 0.297 & NA & NA \\ 
11 & 5h24m59.068s & 69d38m47.891s & 5.977 & 0.555 & -3.097 & 0.288 & 4.924  & -3.693  \\ 
12 & 5h24m58.914s & 69d38m47.797s & 7.006 & 0.578 & -2.873 & 0.237 & 7.386 & -2.462  \\ 
13 & 5h24m59.357s & 69d38m46.738s & 6.259 & 0.589 & -2.238 & 0.210 & NA & NA \\ 
14 & 5h24m59.052s & 69d38m43.732s & 4.547 & 0.578 & -1.877 & 0.238 & 3.693  & -1.231   \\ 
15 & 5h24m58.979s & 69d38m35.721s & 3.913 & 0.571 & 1.736 & 0.253 & NA & NA \\ 
16 & 5h24m58.891s & 69d38m34.856s & 4.518 & 0.513 & 3.136 & 0.356 & 4.924  & 2.462   \\ 
17 & 5h24m59.026s & 69d38m35.230s & 4.951 & 0.505 & 3.605 & 0.368 & NA & NA \\ 
18 & 5h24m59.091s & 69d38m33.413s & 4.812 & 0.543 & 2.746 & 0.310 & 2.462  & 2.462   \\ 
19 & 5h24m59.355s & 69d38m35.500s & 4.221 & 0.427 & 4.511 & 0.456 & NA & NA \\ 
20 & 5h24m59.444s & 69d38m35.482s & 3.312 & 0.478 & 2.791 & 0.403 & NA & NA \\ 
21 & 5h24m59.387s & 69d38m35.145s & 5.867 & 0.534 & 3.571 & 0.325 & NA & NA  \\ 
22 & 5h24m59.471s & 69d38m35.149s & 3.199 & 0.407 & 3.728 & 0.474 & NA & NA  \\ 
23 & 5h24m59.061s & 69d38m32.842s & 3.358 & 0.462 & 3.061 & 0.421 & 2.462  & 2.462  \\ 
24 & 5h24m59.165s & 69d38m32.899s & 6.936 & 0.545 & 3.905 & 0.307 & 4.924  & 3.693  \\ 
25 & 5h24m59.336s & 69d38m33.396s & 5.870 & 0.548 & 3.218 & 0.300 & NA & NA \\ 
26 & 5h25m00.580s & 69d38m33.323s & 1.729 & 0.248 & 4.002 & 0.574 & 1.231  & 1.231  \\ 
27 & 5h25m00.657s & 69d38m33.912s & 1.930 & 0.281 & 3.832 & 0.558 & NA & NA  \\ 
28 & 5h25m00.707s & 69d38m35.283s & 2.645 & 0.433 & 2.750 & 0.451 & NA & NA  \\ 
29 & 5h25m00.614s & 69d38m30.296s & 2.558 & 0.324 & 4.227 & 0.535 & NA & NA  \\ 
30 & 5h25m00.476s & 69d38m33.102s & 1.871 & 0.244 & 4.407 & 0.575 & NA & NA \\ 
31 & 5h25m00.700s & 69d38m28.663s & 0.142 & 0.015 & 5.776 & 0.625 & 0.000 & 3.693 \\ 
32 & 5h25m00.845s & 69d38m27.202s & 2.276 & 0.252 & 5.171 & 0.572 & 3.693  & 2.462 \\ 
33 & 5h25m00.594s & 69d38m26.570s & 1.869 & 0.145 & 7.859 & 0.608 & NA & NA \\ 
34 & 5h25m00.948s & 69d38m26.000s & 2.112 & 0.195 & 6.414 & 0.594 & 3.693 0 & 2.462  \\ 
35 & 5h25m00.447s & 69d38m23.793s & 1.778 & 0.177 & 6.031 & 0.599 & 1.231  & 6.155  \\ 
36 & 5h25m00.407s & 69d38m22.524s & 1.459 & 0.119 & 7.525 & 0.614 & 2.462  & 7.386   \\ 
37 & 5h25m00.231s & 69d38m22.238s & 2.345 & 0.183 & 7.649 & 0.598 & NA & NA  \\ 
38 & 5h25m00.473s & 69d38m23.509s & 3.180 & 0.264 & 6.810 & 0.566 & 4.924  & 7.386   \\ 
39 & 5h24m59.998s & 69d38m22.286s & 2.610 & 0.237 & 6.370 & 0.578 & NA & NA \\ 
40 & 5h25m00.851s & 69d38m18.795s & 1.378 & 0.099 & 8.562 & 0.617 & 1.231 & 6.155  \\ 
41 & 5h25m00.920s & 69d38m18.666s & 2.360 & 0.233 & 5.875 & 0.580 & 0.000 & 6.155   \\ 
42 & 5h25m00.956s & 69d38m17.995s & 0.808 & 0.056 & 8.964 & 0.622 & NA & NA \\ 
43 & 5h25m00.964s & 69d38m18.212s & 1.658 & 0.138 & 7.308 & 0.610 & 0.000  & 6.155   \\ 
44 & 5h25m00.764s & 69d38m18.370s & 1.616 & 0.099 & 10.030 & 0.617 & 2.462 & 12.311   \\ 
45 & 5h25m00.892s & 69d38m18.986s & 1.431 & 0.113 & 7.782 & 0.615 & 1.231 & 6.155   \\ 
46 & 5h25m01.869s & 69d38m20.792s & -1.263 & 0.074 & 10.628 & 0.621 & NA & NA \\ 
47 & 5h25m02.200s & 69d38m20.383s & -1.787 & 0.150 & 7.249 & 0.607 & -3.693 & 4.924   \\ 
48 & 5h25m02.373s & 69d38m20.365s & -2.316 & 0.191 & 7.216 & 0.595 & -3.693  & 4.924  \\ 
49 & 5h25m02.258s & 69d38m20.524s & -2.631 & 0.187 & 8.404 & 0.596 & -3.693  & 4.924   \\ 
50 & 5h25m02.309s & 69d38m21.170s & -1.858 & 0.105 & 10.877 & 0.616 & -2.462  & 11.080 \\ 
51 & 5h25m02.559s & 69d38m26.825s & -3.684 & 0.396 & 4.500 & 0.484 & -3.693  & 3.693   \\ 
52 & 5h25m02.397s & 69d38m30.352s & -1.014 & 0.141 & 4.365 & 0.609 & 0.000  & 4.924   \\ 
53 & 5h25m02.455s & 69d38m30.719s & -1.352 & 0.141 & 5.836 & 0.609 & -1.231  & 6.155   \\ 
54 & 5h25m02.538s & 69d38m30.703s & -4.625 & 0.405 & 5.428 & 0.476 & -3.693  & 4.924   \\ 
55 & 5h25m03.081s & 69d38m30.761s & -2.899 & 0.377 & 3.830 & 0.498 & -1.231 & 3.693   \\ 
56 & 5h25m03.079s & 69d38m29.633s & -2.074 & 0.300 & 3.791 & 0.548 & -2.462 & 4.924   \\ 
57 & 5h25m03.225s & 69d38m29.853s & -4.040 & 0.461 & 3.700 & 0.422 & -1.231  & 2.462   \\ 
58 & 5h25m03.387s & 69d38m28.857s & -2.748 & 0.259 & 6.035 & 0.569 & -3.693  & 4.924   \\ 
59 & 5h25m03.553s & 69d38m31.191s & -4.421 & 0.466 & 3.954 & 0.417 & -3.693  & 3.693  \\ 
60 & 5h25m03.644s & 69d38m31.047s & -4.108 & 0.393 & 5.077 & 0.486 & -4.924  & 2.462  \\ 
61 & 5h25m03.738s & 69d38m30.971s & -4.446 & 0.443 & 4.420 & 0.441 & -2.462  & 3.693  \\ 
62 & 5h25m03.792s & 69d38m30.488s & -3.843 & 0.400 & 4.604 & 0.480 & -2.462 & 4.924   \\ 
63 & 5h25m04.737s & 69d38m37.017s & -4.387 & 0.599 & 1.320 & 0.180 & NA & NA  \\ 
64 & 5h25m04.584s & 69d38m38.620s & -7.044 & 0.540 & 4.111 & 0.315 & NA & NA  \\ 
65 & 5h25m04.800s & 69d38m36.432s & -6.623 & 0.483 & 5.433 & 0.396 & NA & NA  \\ 
66 & 5h25m03.965s & 69d38m41.964s & -2.947 & 0.602 & 0.823 & 0.168 & -2.462  & 1.231  \\ 
67 & 5h25m04.343s & 69d38m40.948s & -4.563 & 0.553 & 2.407 & 0.292 & NA & NA \\ 
68 & 5h25m03.941s & 69d38m42.226s & -5.002 & 0.622 & 0.481 & 0.060 & -2.462  & 1.231  \\ 
69 & 5h25m04.387s & 69d38m30.020s & -5.281 & 0.410 & 6.064 & 0.471 & -6.155  & 2.462  \\ 
70 & 5h25m04.664s & 69d38m29.489s & -5.792 & 0.441 & 5.809 & 0.443 & NA & NA \\ 
71 & 5h25m04.712s & 69d38m29.118s & -5.238 & 0.465 & 4.714 & 0.418 & NA & NA  \\ 
72 & 5h25m05.020s & 69d38m28.243s & -5.438 & 0.469 & 4.799 & 0.414 & -4.924  & 6.155   \\ 
73 & 5h25m05.009s & 69d38m28.027s & -4.951 & 0.417 & 5.517 & 0.465 & -4.924  & 6.155  \\ 
74 & 5h25m04.751s & 69d38m27.468s & -4.284 & 0.342 & 6.548 & 0.523 & -3.693  & 6.155  \\ 
75 & 5h25m05.117s & 69d38m28.383s & -3.508 & 0.360 & 4.986 & 0.511 & -3.693  & 7.386  \\ 
76 & 5h25m04.831s & 69d38m26.683s & -4.528 & 0.413 & 5.136 & 0.469 & -4.924  & 7.386  \\ 
77 & 5h25m04.797s & 69d38m26.002s & -5.496 & 0.402 & 6.537 & 0.478 & -6.155  & 7.386  \\ 
78 & 5h25m02.870s & 69d38m21.258s & -3.230 & 0.273 & 6.654 & 0.562 & -2.462  & 4.924   \\ 
79 & 5h25m02.811s & 69d38m21.684s & -5.987 & 0.404 & 7.064 & 0.477 & -2.462  & 4.924   \\ 
80 & 5h25m03.575s & 69d38m22.023s & -3.531 & 0.270 & 7.370 & 0.564 & 0.000  & 1.231   \\ 
81 & 5h25m03.874s & 69d38m20.501s & -3.165 & 0.291 & 6.015 & 0.553 & -3.693 & 7.386   \\ 
82 & 5h25m03.436s & 69d38m23.201s & -2.373 & 0.198 & 7.098 & 0.593 & NA & NA  \\ 
83 & 5h25m04.893s & 69d38m22.848s & -4.533 & 0.306 & 8.081 & 0.545 & -3.693 & 7.386  \\ 
84 & 5h25m04.797s & 69d38m21.717s & -7.334 & 0.410 & 8.452 & 0.472 & NA & NA  \\ 
85 & 5h25m04.663s & 69d38m20.967s & -7.508 & 0.428 & 7.979 & 0.455 & NA & NA \\ 
86 & 5h25m04.954s & 69d38m18.989s & -5.288 & 0.281 & 10.497 & 0.558 & NA & NA  \\ 
87 & 5h25m05.116s & 69d38m18.782s & -6.096 & 0.363 & 8.557 & 0.509 & NA & NA  \\ 
88 & 5h25m05.411s & 69d38m19.262s & -7.661 & 0.502 & 5.670 & 0.372 & NA & NA  \\ 
89 & 5h25m08.036s & 69d38m20.676s & -11.101 & 0.512 & 7.791 & 0.359 & -12.311 & 4.924  \\ 
90 & 5h25m07.979s & 69d38m20.386s & -10.437 & 0.505 & 7.603 & 0.368 & -12.311  & 4.924   \\ 
91 & 5h25m07.921s & 69d38m20.738s & -8.091 & 0.394 & 9.951 & 0.485 & -9.848  & 6.155   \\ 
92 & 5h25m07.220s & 69d38m16.762s & -9.750 & 0.454 & 9.215 & 0.429 & -8.617  & 8.617   \\ 
93 & 5h25m07.046s & 69d38m17.499s & -10.700 & 0.463 & 9.720 & 0.420 & NA & NA \\ 
94 & 5h25m06.250s & 69d38m19.801s & -8.183 & 0.501 & 6.116 & 0.374 & -8.617  & 6.155   \\ 
95 & 5h25m06.290s & 69d38m19.507s & -8.475 & 0.492 & 6.644 & 0.386 & -8.617  & 6.155   \\ 
96 & 5h25m06.317s & 69d38m17.245s & -6.808 & 0.433 & 7.101 & 0.451 & -7.386  & 7.386   \\ 
97 & 5h25m05.259s & 69d38m16.604s & -6.608 & 0.401 & 7.889 & 0.479 & -7.386 & 7.386   \\ 
98 & 5h25m05.423s & 69d38m15.945s & -5.608 & 0.298 & 10.350 & 0.550 & NA & NA  \\ 
99 & 5h25m05.352s & 69d38m15.983s & -6.149 & 0.357 & 8.852 & 0.513 & NA & NA  \\ 
100 & 5h25m05.648s & 69d38m15.853s & -6.412 & 0.367 & 8.849 & 0.506 & -6.155  & 8.617  \\ 
101 & 5h25m05.767s & 69d38m16.143s & -6.894 & 0.343 & 10.506 & 0.523 & -6.155 & 8.617   \\ 
102 & 5h25m05.794s & 69d38m15.290s & -6.897 & 0.366 & 9.563 & 0.507 & NA & NA  \\ 
103 & 5h25m06.181s & 69d38m13.869s & -7.882 & 0.410 & 9.053 & 0.471 & -9.848 & 8.617  \\ 
104 & 5h25m05.608s & 69d38m13.780s & -6.183 & 0.314 & 10.619 & 0.540 & -6.155  & 8.617 \\ 
105 & 5h25m06.015s & 69d38m13.013s & -8.498 & 0.391 & 10.610 & 0.488 & NA & NA \\ 
106 & 5h25m05.661s & 69d38m13.407s & -6.182 & 0.323 & 10.241 & 0.535 & -6.155  & 8.617  \\ 
107 & 5h25m05.593s & 69d38m12.558s & -6.322 & 0.313 & 10.925 & 0.541 & NA & NA  \\ 
108 & 5h25m07.099s & 69d38m11.216s & -9.856 & 0.435 & 10.153 & 0.448 & -9.848  & 9.848  \\ 
109 & 5h25m06.241s & 69d38m10.852s & -7.146 & 0.375 & 9.535 & 0.500 & -7.386  & 8.617  \\ 
110 & 5h25m05.609s & 69d38m10.424s & -7.850 & 0.387 & 9.943 & 0.491 & -6.155  & 9.848   \\ 
111 & 5h25m05.343s & 69d38m08.948s & -5.298 & 0.301 & 9.658 & 0.548 & NA & NA  \\ 
112 & 5h25m05.381s & 69d38m11.992s & -6.841 & 0.346 & 10.297 & 0.521 & NA & NA \\ 
113 & 5h25m05.438s & 69d38m10.191s & -7.160 & 0.344 & 10.850 & 0.522 & -6.155 & 9.848  \\ 
114 & 5h25m04.640s & 69d38m10.040s & -3.924 & 0.241 & 9.388 & 0.577 & -7.386  & 9.848   \\ 
115 & 5h25m04.454s & 69d38m09.844s & -4.261 & 0.246 & 9.951 & 0.575 & NA & NA  \\ 
116 & 5h24m58.646s & 69d38m25.981s & 6.982 & 0.425 & 7.526 & 0.458 & NA & NA \\ 
117 & 5h25m01.395s & 69d38m49.800s & 1.921 & 0.302 & -3.487 & 0.547 & NA & NA  \\ 
118 & 5h25m01.442s & 69d38m50.223s & 0.035 & 0.005 & -4.109 & 0.625 & NA & NA  \\ 
119 & 5h24m59.732s & 69d38m20.293s & 3.725 & 0.301 & 6.767 & 0.548 & 0.000 & 3.693  \\ 
120 & 5h25m05.432s & 69d38m08.588s & -6.583 & 0.334 & 10.405 & 0.528 & NA & NA  
\enddata
\tablecomments{Positive values indicate direction to the north and east for RA and Dec. respectively. Results of the automated procedure are matched when applicable.}
\end{deluxetable*}

\hfill \break

\end{document}